\documentclass[a4,12pt]{article}
\usepackage{type1cm,amsmath,amssymb,color,graphics,amscd,amsfonts,epsf,indentfirst,cite}
\usepackage{epsfig}
\usepackage{subfigure}
\usepackage{bm}
\usepackage{ulem}

\newcommand{\calL}{\mathcal{L}}
\newcommand{\calT}{\mathcal{T}}
\newcommand{\bgamma}{\boldsymbol{\gamma}}
\newcommand{\bp}{\boldsymbol{p}}

\newcommand{\bA}{\boldsymbol{A}}

\newcommand{\feyn}[1]{
  \setbox0=\hbox{\ensuremath{#1}}
  \hbox to\wd0{\hbox to0pt{\hbox to\wd0{\hss/\hss}\hss}\box0}}

\allowdisplaybreaks[4]

\makeatletter
\@addtoreset{equation}{section}

\makeatletter
\renewcommand\section{\@startsection {section}{1}{\z@}%
                                   {-3.5ex \@plus -1ex \@minus -.2ex}
                                   {2.3ex \@plus.2ex}%
                                   {\normalfont\large\bfseries}}
\renewcommand\subsection{\@startsection{subsection}{2}{\z@}%
                                     {-3.25ex\@plus -1ex \@minus -.2ex}%
                                     {1.5ex \@plus .2ex}%
                                    {\normalfont\bfseries}}

\begin{document}

\begin{titlepage}
  \thispagestyle{empty}

\begin{flushright}
 {\tt 
 OU-HET-913
 }
\\
\end{flushright}
    
  \vspace{0.5cm}
  
  \begin{center}
    \font\titlerm=cmr10 scaled\magstep4
    \font\titlei=cmmi10 scaled\magstep4
    \font\titleis=cmmi7 scaled\magstep4

     \centerline{\titlerm
Holographic Floquet states:}

\vspace{2mm}

     \centerline{\titlerm
(I) A strongly coupled Weyl semimetal}
    
    \vspace{0.6cm}
    \noindent{
   Koji Hashimoto\footnote[1]{koji@phys.sci.osaka-u.ac.jp},
   Shunichiro Kinoshita\footnote[2]{kinoshita@phys.chuo-u.ac.jp},
   Keiju Murata\footnote[3]{keiju@phys-h.keio.ac.jp} and
   Takashi Oka\footnote[4]{oka@pks.mpg.de}
      }\\
    \vspace{0.5cm}
   $^{1}${\it Department of Physics, Osaka University,
 Toyonaka, Osaka 560-0043, Japan}
   \\
   $^{2}${\it Department of Physics, Chuo University, Tokyo 112-8551, Japan}
   \\
   $^{3}${\it Keio University, 4-1-1 Hiyoshi, Yokohama 223-8521, Japan}
   \\
   $^{4}${\it Max-Planck-Institut f\"{u}r Physik komplexer Systeme (MPI-PKS), 
   N\"{o}thnitzer Stra\ss e 38, Dresden 01187, Germany}
   \\
   $^{4}$ {\it Max-Planck-Institut f\"{u}r Chemische Physik fester Stoffe (MPI-CPfS), 
   N\"{o}thnitzer Stra\ss e 40, Dresden 01187, Germany
   }

  \end{center}  
  
  \vskip 1em

\begin{abstract}
Floquet states can be realized in quantum systems driven by continuous time-periodic perturbations. 
It is known that a state known as the Floquet Weyl semimetal can be realized when 
free Dirac fermions are placed in a rotating electric field. 
What will happen if strong interaction is introduced to this system? 
Will the interaction wash out  the characteristic features of Weyl semimetals such as the Hall response? 
Is there a steady state and what is its thermodynamic behavior? 
We answer these questions using AdS/CFT correspondence in
the ${\cal N}=2$ supersymmetric massless QCD in a rotating electric field in the large $N_c$ limit
realizing the first example of a ``holographic Floquet state''. 
 In this limit, gluons not only mediate interaction, but also act as an energy reservoir and stabilize the nonequilibrium steady state (NESS). 
We obtain the electric current induced by a rotating electric field:
In the high frequency region, the Ohm's law is satisfied, while
we recover the DC nonlinear conductivity at low frequency, 
which was obtained holographically in a previous work. 
The thermodynamic properties of the NESS, e.g.,  fluctuation-dissipation relation, 
is characterized by the effective Hawking temperature 
that is defined from the  effective horizon giving a holographic 
meaning to the ``periodic thermodynamic'' concept. 
In addition to the strong (pump) rotating electric field, 
we apply an additional weak (probe) electric field in the spirit of the 
pump-probe experiments done in condensed matter experiments. 
Weak DC and AC probe analysis in the background rotating electric field
shows Hall currents as a linear response, therefore the Hall response
of Floquet Weyl semimetals survives at the strong coupling limit.
We also find frequency mixed response currents, i.e.,  a heterodyning effect, 
characteristic to periodically driven Floquet systems.

\end{abstract}

\end{titlepage}


\tableofcontents

\clearpage

\section{Introduction}

\begin{figure*}[htb]
\begin{center}
\includegraphics[scale=1.0]{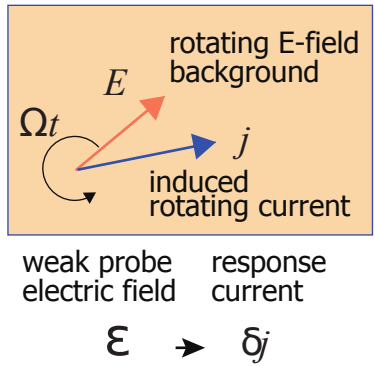}
\end{center}
 \caption{
Setup for the holographic calculation. In the steady state, the rotating electric field $E$ within the $(x,y)$-plane 
induces a retarded rotating current $j$ in the $(x,y)$-plane. In addition to $E$, 
 a weak non-rotating electric field $\varepsilon$ is applied. This field induces a current $\delta j$ linear in $\varepsilon$.
}
 \label{setup}
\end{figure*}

\begin{figure*}[htb]
\centering 
\includegraphics[width=12cm]{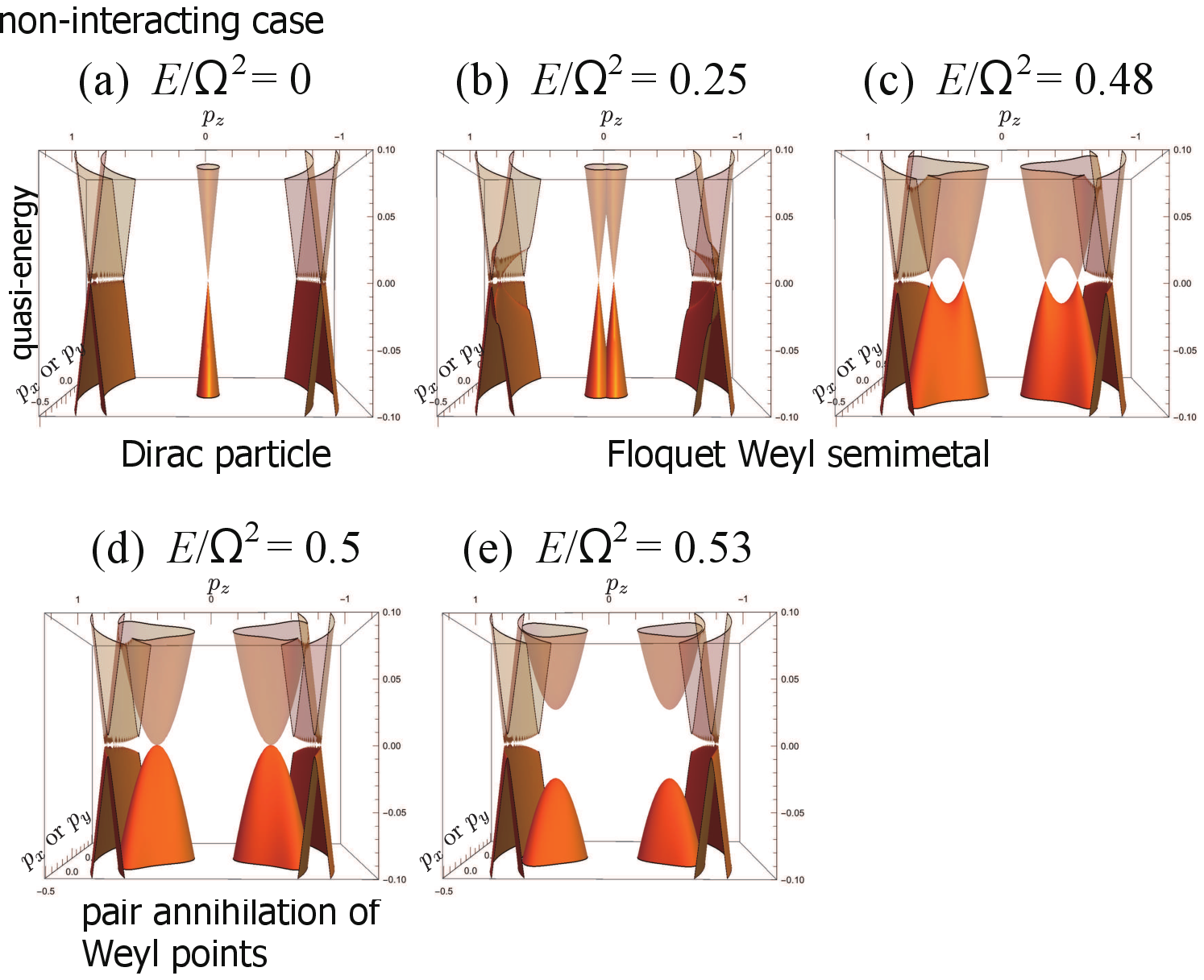}
\caption{
 Quasi-energy spectrum of a noninteracting Dirac particle in rotating electric fields,
 taken from \cite{Oka-new}.
(a) At zero field, spectrum of massless Dirac particles exists at $\vec{p}=0$ and the Floquet sidebands 
for states with energies $\pm |\vec{p}|+m \Omega$ surrounds it ($m$: integer). 
(b,c) For fields $0<E/\Omega^2\le 1/2$, the Dirac spectrum splits into spectra of two Weyl fermions (\ref{eq:Leff}). 
At $p_z=\pm 1, p_x=p_y=0$, a new sets of Weyl points emerge and move toward the two Weyl points near the center. 
(d) The central Weyl points pair annihilates with the emergent Weyl points  at fields $E/\Omega^2=1/2$. 
(e) A gap opens at $p_z=\pm 1/2,p_x=p_y=0$. 
}
\label{fig:FloquetSpectrum}
\end{figure*}

Nonequilibrium phenomenon in strongly correlated systems is of general interest
and the machinery of the renowned AdS/CFT correspondence
\cite{Maldacena:1997re,Gubser:1998bc,Witten:1998qj} can assist us reveal its nature
(Its application to equilibrium condensed matter problems can be found in refs.~\cite{Hartnoll:2008,Cubrovic:2009,Huijse:2012}).
In this article, we apply it to study the massless QCD in a rotating electric field 
\begin{equation}
\vec{E}(t)=
  \begin{pmatrix}
    \cos\Omega t \\
  \sin\Omega t 
  \end{pmatrix}
E
  \ ,\qquad (E\geq 0)
  \label{rotEbc}
\end{equation}
where the field with constant strength $E$ 
is rotating anticlockwise with a frequency $\Omega$ in the $(x,y)$-plane. The schematic picture of the setup is shown in Fig.~\ref{setup}.
This background field drives the system away from equilibrium and
generates a nonequilibrium state with broken time reversal symmetry. 
When a free Dirac particle in $(3+1)$ dimensions 
with a Lagrangian 
$\calL= \bar{\psi}(i\feyn{\partial}-e\feyn{A})\psi$  
($\vec{E}=-\partial_t\vec{A}$,
$\feyn{A}=\gamma^\mu A_\mu$ and $\gamma^\mu$ are gamma matrices)
is considered, the field will dynamically generate a constant axial vector field in the $z$-direction. 
The resulting static effective Lagrangian is given by 
\begin{equation}
  \calL_{\text{eff}} = \bar{\psi}i\feyn{\partial}\psi
  - b\bar{\psi}\gamma_5\gamma^z\psi 
  \label{eq:Leff}
\end{equation}
with $b=(eE)^2/\Omega^3+\mathcal{O}(1/\Omega^5)$ \cite{Ebihara:2016}.
Floquet theorem is a temporal analogue of the Bloch theorem and 
enables us to systematically study periodically driven states by mapping 
it to a time-independent effective problem with the aid of Fourier mode expansion \cite{Sambe:1973}. 
Using proper driving fields, it is possible to control the topological properties of the 
system since we can induce adequate terms necessary to realize non-trivial topology
in the spectrum as in the axial vector field $b$ in (\ref{eq:Leff}). 
The study of ``Floquet topological insulators''~\cite{Lindner11,Oka09,Kitagawa10,Kitagawa11}
is becoming a hot topic in condensed matter systems 
and recently Haldane's topological lattice model~\cite{Haldane1988} was experimentally realized~\cite{Jotzu14} by 
applying  a rotating electric field to fermions in a honeycomb lattice \cite{Oka09}. 
In $(2+1)$ dimensions, 
the free massless Dirac particle shows a gap opening in rotating electric fields 
leading to an emergent parity anomaly (Hall effect), where the direction of the 
Hall current, i.e., a current flowing perpendicular to the applied DC-electric field, can be 
controlled by changing the polarization of the field. 
The gap opening in $(2+1)$-dimensional 
Dirac fermions was already experimentally observed 
in an ultrafast pump-probe experiment \cite{Wang13}. 
Now, moving on to a $(3+1)$-dimensional 
Dirac system, an analogous phenomena take place and it was predicted that
the Dirac point splits into two Weyl points \cite{Nielsen:1983} forming a 
``Floquet Weyl semimetal'' ~\cite{Wang:2014} with broken time reversal symmetry
followed by refs.~\cite{Ebihara:2016,XiaoXiao:2016,Hubener:2016,Oka-new}. 
 Figure~\ref{fig:FloquetSpectrum}(a-e) shows the numerically exact Floquet quasienergy and 
 we clearly see the two Weyl points in (b) (see section \ref{sec:weakcoupling} for further details).

At least in the weak coupling limit,
Floquet Weyl semimetals can be created from Dirac semimetals by applying the rotating electric field.
Here, we arise following questions. 
{\it{What will happen in a strongly interacting fermion systems? 
Will the interaction wash out the characteristic features of Weyl semimetals such as Hall response? 
Is there a nonequilibrium steady state and what is the thermodynamic behavior?}}
We address these questions using AdS/CFT correspondence.
``Periodic thermodynamics'' is another important topic in the research of periodically driven 
many-body systems \cite{Kohn:2014, Lazarides:2014,Lazarides:20142,DAlessio:2014}.  
Whether or not thermodynamic concepts 
such as the variational principle, temperature and universal distributions, e.g., Gibbs ensemble, 
still has a meaning in driven systems is an 
interesting question. 
To realize ``periodic thermodynamics'',  it is necessary to 
stabilize the system because it is constantly heated up by the driving. 
One way to fulfill this is to attach the system to a heat reservoir \cite{Kohn:2014}, while  
Floquet topological states coupled to a boson bath has been studied in Refs.~\cite{Dehghani:2014,Dehghani:20152}. 

AdS/CFT correspondence can provide a framework to study these problems
in a strongly coupled field theory. 
As a toy model, we focus on $\mathcal{N}=2$ $SU(N_c)$ supersymmetric QCD at large $N_c$ and 't Hooft coupling. 
This theory is geometrically realized as a probe D7-brane in AdS$_5\times S^5$ spacetime~\cite{Karch}.
The low energy theory has a single quark field (D7-brane) and $SU(N_c)$ gluon fields (AdS$_5\times S^5$ background) that mediates interaction.
In order to mimic the Floquet Weyl semimetal, we set the quark to be massless and apply the external rotating electric field. 
Since there are infinite number of gluon fields in the large $N_c$ limit, the gluons not only mediate a 
long range interaction between the quarks, but also acts as a thermal reservoir. Thus, a nonequilibrium steady state (NESS) may be 
stabilized even if the system is constantly heated up by external driving.

We will demonstrate first analysis of the nonlinear response against
the strong and rotating electric field applied 
to $\mathcal{N}=2$ supersymmetric QCD via the AdS/CFT correspondence.
Let us briefly look at relevant history of calculations of nonlinear
and linear responses against electric field
in gravity models
to illustrate our analysis. 
A strong and constant electric field in D3-D7 model was first considered in 
Refs.~\cite{Karch:2007pd,Albash:2007bq,Erdmenger:2007bn}.
The quark electric current $j$ was found to be proportional to $E^{3/2}$
for the case of massless supersymmetric QCD, as a consequence of a 
scale invariance of the theory.
For the case of massive QCD, the nonlinear regime is indispensable also for a ``deconfinement''
transition~\cite{Albash:2007bq,Erdmenger:2007bn,Hashimoto:2013mua,Hashimoto:2014dza,Hashimoto:2014yya}.
(See also Refs.~\cite{Hashimoto:2014yza,Hashimoto:2014xta,Hashimoto:2014dda}
for the dynamical phase transition by the electric field quench in the D3-D7 model.) 
The probe AC conductivities have also been widely studied 
in AdS/CFT especially in its application to condensed matter physics.
For example, in the holographic superconductor~\cite{Hartnoll:2008kx,Hartnoll:2008vx},
AC conductivity under a weak AC electric field was calculated, and a superconductivity gap
was explicitly obtained. (See Refs.\cite{Hartnoll:2009sz,Herzog:2009xv,Sachdev:2010ch} for comprehensive reviews on the AC conductivities in various gravity models using AdS/CFT.)
The nonlinear response against strong AC electric field in the holographic superconductor
is also studied in Refs.\cite{Li:2013fhw,Natsuume:2013lfa}.
For the strong and time-dependent electric field, its response
has not been studied so much because of its technical difficulty:
We need to solve nonlinear partial differential equations obtained in the gravity side.
We overcome the difficulty by considering the rotating electric field.
We will see that the resultant equations of motion reduce to ordinary
differential equations for the rotating electric field even in the nonlinear regime.

This paper is organized as follows.
First, in section 2, we shall introduce
a free fermion picture of
Weyl semimetal under a rotating electric field, using Floquet method.
Then in section 3,
we describe the gravity dual of strongly coupled ${\cal N}=2$ supersymmetric massless QCD at large $N_c$,
and introduce the rotating external electric field.
We obtain the nonlinear conductivity explicitly as a function of the intensity of
the  electric field and the frequency $\Omega$.
In particular, in the DC limit $\Omega \to 0$, our result reproduces the
$j\propto E^{3/2}$ law found above, while in the other limit $\Omega \to \infty$ we find
that the system is subject to the Ohm's law.
In section 4,
we calculate linear DC and AC response to the background rotating
electric field, and discover a holographic Hall effect: Electric currents can have a component 
perpendicular to the introduced (probe) weak DC and AC electric fields.
We also find frequency mixed response currents characteristic to periodically driven Floquet systems. 
Appendices are given for explicit notations and
numerical recipes.

\section{Floquet Weyl semimetal at weak coupling}
\label{sec:weakcoupling}

\label{sec:noninteracting}
Here, we describe the Floquet spectrum of a free massless Dirac 
particle in a rotating electric field \cite{Wang:2014,Ebihara:2016}. 
In the rotating electric field the Hamiltonian $H(t)$ becomes
periodically time-dependent, i.e., $H(t+T)=H(t)$ where $T=2\pi/\Omega$ is the periodicity. 
Quantum states in time periodic driving are described by the Floquet
theory~\cite{Sambe:1973,Shirley:1965}, that is, a temporal version of
the Bloch theorem.  The essence of the Floquet theory is a mapping
between the time-dependent Schr\"odinger equation and a static
eigenvalue problem.  The eigenvalue is called the Floquet
pseudo-energy and plays a role similar to the energy in a static system.  

Let us consider a Hamiltonian, $H_{\rm tot}=H_0+H_{\rm bg}$, with
\begin{equation}
  H_0 = \gamma^0\bgamma\cdot\bp + \gamma^0 m\;, \qquad
  H_{\rm bg} = -e\gamma^0\bgamma\cdot\bA\;,
\end{equation}
that describes the one-particle Dirac system coupled to an external background 
gauge field and $\gamma^\mu$ are the Dirac matrices satisfying
$\{\gamma^\mu,\gamma^\nu\}=2\eta^{\mu\nu}$.  In an rotating electric field
in the $(x,y)$-plane (\ref{rotEbc}), we can write the
time-dependent vector potential as
\begin{equation}
  A_x = -\frac{E}{\Omega}\sin(\Omega t)\;,\quad
  A_y = \frac{E}{\Omega}\cos(\Omega t)\;,\quad
  A_z=0\;,
\end{equation}
where $\Omega$ is the frequency.  We can conveniently decompose the
interaction part of the Hamiltonian into two pieces as
$H_{\rm bg}=e^{i\Omega t}H_- + e^{-i\Omega t}H_+$ where
$H_\pm=\pm i(eE/\Omega)\gamma^0\gamma^{\pm}$ with
$\gamma^\pm=\frac{1}{2}(\gamma^x\pm i\gamma^y)$.  
The Floquet spectrum is defined as the spectrum of the
Hermitian operator  $\mathcal{H}=H_{\rm tot}-i\partial_t$ 
acting on the space of time periodic Floquet states $\Phi(t)$ 
satisfying $\Phi(t+T)=\Phi(t)$.

Now we assume that the the period $T=2\pi/\Omega$ of the circular polarization is small
enough as compared to the typical observation timescale.  We can then
expand the theory in terms of $\omega/\Omega$ (with $\omega$ being a
frequency corresponding to some excitation energy).  Taking the
average over $T$ we can readily find the following effective
Hamiltonian by the Floquet Magnus expansion:
\begin{equation}
  H_{\text{eff}} = \frac{i}{T}\ln\Bigl[ \calT
    e^{-i\int_0^T dt\,H(t)} \Bigr]
  \simeq H_0 + \frac{1}{\Omega}[H_-,H_+] \;,
\end{equation}
to the first order in the expansion \cite{Kitagawa11}.  
Interestingly we can express the induced term as
\begin{equation}
 H_{\text{ind}} \equiv \frac{1}{\Omega}[H_-,H_+]
  = -\frac{(eE)^2}{\Omega^3}i\gamma^x\gamma^y
  = \beta \gamma^0 \gamma_5\gamma^z \;,
\label{eq:Hind}
\end{equation}
where we defined $\beta\equiv(eE)^2/\Omega^3$.  This means that the
circular polarized electric field would induce an axial-vector
background field $\bA_5=\beta\hat{\boldsymbol{z}}$ perpendicular to
the polarization plane \cite{Ebihara:2016}.  A related expression was obtained in
Ref.~\cite{Wang:2014} in the context of ``Floquet Weyl semimetal.''

The effect of finite $\beta$ is easily understandable from the energy
dispersion relations.  We can immediately diagonalize $H_{\text{eff}}$
and the four pseudo-energies read:
\begin{equation}
  \varepsilon_\pm(p)
  = \sqrt{p_x^2 + p_y^2 + (\sqrt{p_z^2+m^2}\pm\beta)^2}
\end{equation}
and $-\varepsilon_\pm(p)$.  
It
shows that the Dirac point splits into
two Weyl points with a displacement given by
\begin{equation}
  \Delta p = \sqrt{\beta^2-m^2} \;.
\end{equation}
In fact, $\beta$ is nothing but a momentum shift along the $z$-axis
that is positive for the right chirality state (i.e.,
$\gamma_5\psi_{\rm R}=+\psi_{\rm R}$) and negative for the left
chirality state (i.e., $\gamma_5\psi_{\rm L}=-\psi_{\rm L}$).
We point out that time- and angle-resolved photoemission spectroscopy 
should be able to 
see this splitting of Weyl points in a similar manner as 
the gap opening~\cite{Oka09} of the $(2+1)$-dimensional 
Dirac point already observed experimentally~\cite{Wang13}.
Interestingly, as long as $\beta>m$, the pseudo-energy always has two
Weyl points (if they are inside of the Brillouin zone) even for $m>0$.
Therefore, we do not have to require strict massless-ness to realize
gapless dispersions, which should be a quite useful feature for
practical applications including the Schwinger or Landau-Zener effect.

The full Floquet spectrum for $m=0$ is obtained numerically by diagonalizing the 
Floquet Hamiltonian \cite{Oka-new}. 
The spectrum is given in Figs.~\ref{fig:FloquetSpectrum}.
Here we briefly describe and summarize features of the spectrum:
\begin{description}
\item[(a) $E/\Omega^2=0$:] 
The Dirac point exists at $\Vec{p}=0$. The degeneracy of this point is 4.
\item[(b)(c) $0<E/\Omega^2<0.5$:] 
Four Weyl points exists. Two Weyl points came from the initial Dirac point
while the other 2 comes from the hybridized state between the 1 photon
absorbed and emitted states. 
\item[(d) $E/\Omega^2=0.5$:]
The two Weyl points vanish
through pair annihilation with the emergent Weyl points originating from the Floquet side bands.\footnote{
Pair annihilation of Weyl points with opposite chiralities is one way to deform the Weyl system. 
Another mechanism is interaction: It was proposed that forward scattering can open a 
gap in the Weyl spectrum~\cite{Morimoto:2016}.}
Two parabolic Dirac points appear at $p_x=p_y=0,\; p_z=\pm 0.5$. 
\item[(e) $E/\Omega^2=0.5$:] 
Gap opens at the Dirac points. 
\end{description}
The full result is consistent with the perturbative result. Applying the rotating electric field
shifts the location of the Weyl points, and the Dirac point is separated into two Weyl nodes.

\section{Floquet  state in AdS/CFT}

\label{sec:FWS}
\subsection{Set up}

In the previous section, we have argued that, in the weak coupling limit, the Floquet Weyl semimetal can be created from the Dirac semimetal by a rotating electric field.
Here, we consider a similar set up in the strong coupling limit using AdS/CFT correspondence.
As a toy model of the strongly coupled field theory,
we focus on $\mathcal{N}=2$ $SU(N_c)$ supersymmetric QCD at large $N_c$ and at strong 't Hooft coupling.
Using AdS/CFT correspondence, $\mathcal{N}=2$ SQCD is realized by the probe D7-brane in AdS$_5\times S^5$ spacetime~\cite{Karch}.
We expect that quarks and antiquarks in this theory play the roles of
electrons and holes in condensed matter systems.
Applying the rotating external electric field in the system, we will
realize the gravity dual of a Floquet state in $\mathcal{N}=2$ SQCD.
We use the following coordinates for the AdS$_5\times S^5$ spacetime as
\begin{multline}
 ds^2=\frac{\rho^2+w_1^2+w_2^2}{R^2}[-dt^2+dx^2+dy^2+dz^2]\\
  +\frac{R^2}{\rho^2+w_1^2+w_2}[d\rho^2 + \rho^2 d\Omega_3^2 + dw_1^2 +dw_2^2]\ ,
  \label{AdS5S5}
\end{multline}
where $R$ is the AdS radius.
Hereafter, we will take a unit of $R=1$ to simplify the following expressions.

Dynamics of the D7-brane is described by the Dirac-Born-Infeld (DBI) action,
\begin{equation}
 S=-T_7 \int d^8\sigma \sqrt{-\textrm{det}[h_{ab}+2\pi\alpha' F_{ab}]}\ ,
\end{equation}
where $T_7$ is the tension of the brane,
$h_{ab}$ is the induced metric and $F_{ab}=\partial_a A_b-\partial_b A_a$ is the $U(1)$-gauge field strength on the brane.
In the AdS$_5\times S^5$ spacetime, $(t,x,y,z,\rho,\Omega_3)$-directions are filled with the D7-brane.
For simplicity, we consider SQCD with massless quarks in the
boundary theory.
The brane configuration corresponding to it is given by a trivial
solution for the brane position: $w_1=w_2=0$. 
The induced metric on the D7-brane is simply written as
\begin{equation}
h_{ab}d\sigma^a d\sigma^b=\rho^2[-dt^2+dx^2+dy^2+dz^2]
 +\frac{1}{\rho^2}[d\rho^2 + \rho^2 d\Omega_3^2]\ .
 \label{indmet}
\end{equation}
Thus, 
the dynamics of the D7-brane is described only by the gauge field $A_a$.
We assume the spherical symmetry of $S^3$ and translational symmetry in $(x,y,z)$-space.
We also set $A_t=A_\rho=A_z=0$, for simplicity, so that there is no baryon
number density in the boundary theory.
Then, the gauge field $A_a$ is written as
\begin{equation}
 (2\pi \alpha') A_a d\sigma^a = a_x(t,\rho) dx + a_y(t,\rho) dy\ .
  \label{arealdef}
\end{equation}
Near the AdS boundary, the rescaled gauge field $\vec{a}=(a_x,a_y)$ is expanded as
\begin{equation}
 \vec{a}(t,\rho)=-\int^t dt' \vec{E}(t') + \frac{\vec{j}(t)}{2\rho^2} 
  +\frac{\dot{\vec{E}}(t)}{2\rho^2} \ln \left(\frac{\rho}{\rho_0}\right) + \cdots\ ,\quad (\rho\to \infty)\ ,
\label{ainf}
\end{equation}
where $\rho_0$ is a constant introduced to non-dimensionalize the argument of the logarithmic term.
The dot denotes $t$-derivative.
The functions $\vec{E}(t)$ and $\vec{j}(t)$ in the series expansion correspond to
the electric field $\vec{\mathcal{E}}(t)$ and current $\vec{J}(t)$ in
the boundary theory
as\footnote{
Note that there is ambiguity in the definition of the electric current
$\vec{J}$ when we consider time-dependent electric fields.
In Eq.~(\ref{ainf}),
$\rho$ is normalized by $\rho_0$ in the logarithmic term.
If we choose a different normalization such as $\ln(\rho/\tilde\rho_0)$, the definition of $\vec{j}$ is changed as
$\vec{j}\to \vec{j}+\dot{\vec{E}} \ln(\tilde\rho_0/\rho_0)$.
This ambiguity originates from the ambiguity in the finite local counter term in the DBI action, which should be fixed by a
renormalization condition.
}
\begin{equation}
\vec{\mathcal{E}}(t)=\left(\frac{\lambda}{2\pi^2}\right)^{1/2}\vec{E}(t)
\ ,\qquad
\vec{J}(t)= \frac{N_c\sqrt{\lambda}}{2^{5/2}\pi^3}\,\vec{j}(t)
\ .
\end{equation}

In this paper, we will consider a rotating external electric field as in Eq.(\ref{rotEbc}).
The electric field with a constant strength $E$ is parallel to $x$-axis at the initial time $t=0$ and
rotating anticlockwise with a frequency $\Omega$ in $(x,y)$-plane.
To provide such rotating electric fields conveniently, 
we introduce the following complex variable, 
\begin{equation}
 a(t,\rho)\equiv a_x(t,\rho)+ia_y(t,\rho)\ .
  \label{coma}
\end{equation}
Using Eqs.~(\ref{indmet}), (\ref{arealdef}) and (\ref{coma}), 
we obtain the action for $a(t,\rho)$ as
\begin{equation}
S= -T_7 \Omega_3 V_3 \int dt d\rho \, \rho
\big[
\rho^4
-|\dot{a}|^2
+\rho^4|a'|^2
-\{\textrm{Im}(\dot{a}a'{}^\ast)\}^2
\big]^{1/2}\ ,
\label{action_a}
\end{equation}
where $\Omega_3=\textrm{Vol}(S^3)$ and $V_3=\int dxdydz$.
For the complex field $a$, the boundary condition~(\ref{rotEbc})
is written as
\begin{equation}
 a|_{\rho=\infty}=\frac{iE}{\Omega} e^{i\Omega t}\ .
  \label{bc_a}
\end{equation}
Now, we define a new complex variable $b(t,\rho)$
by factoring out a time-dependent phase factor
as
\begin{equation}
 a(t,\rho)=e^{i\Omega t} b(t,\rho)\ .
\label{bdef}
\end{equation}
For this variable $b$, the above boundary condition becomes time-independent as
\begin{equation}
 b|_{\rho=\infty}=\frac{iE}{\Omega}\ .
  \label{bc_b}
\end{equation}
Note that
$E$ may be a complex constant in general, and its magnitude and phase
describe the strength of the electric field and the alignment of the
electric field at $t=0$, respectively.

The action for $b$ is written as 
\begin{equation}
 S= -T_7 \Omega_3 V_3 \int dt d\rho \,
 \rho
\big[
\rho^4
-|(\partial_t+i\Omega)b|^2
+\rho^4|b'|^2
-\{\textrm{Im}((\partial_t+i\Omega)b \, b'{}^\ast)\}^2
 \big]^{1/2}\ .
 \label{action_b}
\end{equation}
In addition to
the time-independent boundary condition~(\ref{bc_b}),
this action does not depend on $t$ explicitly.
Thus, we can consistently assume that the variable $b$ does not depend on $t$:
$b(t,\rho)=b(\rho)$.
Then, the DBI action becomes
\begin{equation}
 S= -T_7 V_4  \int d\rho \mathcal{L}_0\ ,\quad
\mathcal{L}_0\equiv 
\rho
\big[
\rho^4
-\Omega^2|b|^2
+\rho^4|b'|^2
-\Omega^2\{\textrm{Re}(b  b'{}^\ast)\}^2
\big]^{1/2}\ ,
\label{DBI_static}
\end{equation}
where $V_4=\int dtdxdydz$.
This action is invariant under the constant phase rotation: 
$b\to e^{i\alpha}b$, where $\alpha$ is an arbitrary real constant.
Its Noether charge is given by
\begin{equation}
 q = \frac{\Omega \rho^6}{\mathcal{L}_0}\textrm{Im}(b'b^\ast)\ .
\label{qjule}
\end{equation}
This quantity is conserved along the $\rho$-direction: $dq/d\rho=0$.
We will see later that the conserved charge $q$ corresponds to the Joule
heating in the boundary theory.
The equation of motion for $b$ is given by
\begin{multline}
 b''=\frac{1}{4\rho(\rho^4-\Omega^2|b|^2)}\big[
-12\rho^4 b'
-4\rho (\Omega^2 b  +  3\rho^3 b'{}^2 b'{}^\ast)\\
+\Omega^2 \{4b^\ast  b'  (3b-\rho b')-8b'{}^\ast b^2\}
+\Omega^2 b'(bb^\ast)'{}^2
\big]\ .
\label{beq}
\end{multline}
We can obtain the equation of motion for $b^\ast$ by taking the
complex conjugate of the above equation.
It is remarkable that equations of motion for the brane dynamics
reduce to ordinary differential equations (ODEs)
even though the electric field is time-dependent.
This dramatically simplifies our following analysis.
For the reduction to ODEs, it is essential to consider the 
the rotating electric field~(\ref{rotEbc}).
If we consider a linearly (or elliptically) polarized field such as
$\vec{E}=E (\cos\Omega t, 0)$ instead,
the equations of motion will be given by 
partial differential equations of $t$ and $\rho$.

\subsection{Boundary conditions at the effective horizon and the AdS boundary}

The equation of motion~(\ref{beq}) is singular at $\rho=\rho_c$ where $\rho_c$ satisfies 
\begin{equation}
 b(\rho_c)=\frac{\rho_c^2}{\Omega}e^{i\theta}\ ,
\label{brhoc}
\end{equation}
where $\theta$ is a real constant.
In section~\ref{sec:Effmet}, we will see that the singular surface $\rho=\rho_c$ is an
effective event horizon in an effective geometry of the D7-brane.
We impose regularity on the solution $b(\rho)$ at $\rho=\rho_c$ for
physical quantities to be regular.
Expanding the 
solution around $\rho=\rho_c$, we obtain the series expansion of the regular solution as
\begin{equation}
 b(\rho)=e^{i\theta}\frac{\rho_c^2}{\Omega}
+e^{i\theta} p\,
(\rho-\rho_c) + \cdots\ ,\quad
p\equiv 
\mathcal{R}_6-\sqrt{\mathcal{R}_4\mathcal{R}_9}+i\sqrt{2\mathcal{R}_4(\sqrt{\mathcal{R}_4\mathcal{R}_9}-\mathcal{R}_6)}
\label{horreg}
\end{equation}
where we define $\mathcal{R}_n=1/\bar{\rho}_c+n\bar{\rho}_c$
and $\bar{\rho}_c=\rho_c/\Omega$.
The derivation of the asymptotic solution is summarized in appendix~\ref{app:regsol}.
Substituting the above expression into Eq.~(\ref{qjule}), we can simply write the
conserved charge as
\begin{equation}
 q=\rho_c^5\ .
\label{qofx}
\end{equation}

On the other hand, near the infinity, the asymptotic solution becomes\footnote{
We will normalize $\rho$ by $\rho_0=\Omega$ in the logarithmic term throughout this paper.
}
\begin{equation}
 b(\rho)=\frac{iE}{\Omega}+\frac{j}{2\rho^2}+\frac{i\Omega
  E}{2\rho^2}\ln \left(\frac{\rho}{\Omega}\right) +\cdots\ ,
\label{binf}
\end{equation}
where $E$ and $j$ are complex constants.
Substituting the above expression into Eq.~(\ref{qjule}), we have
a different expression for the 
conserved charge as
\begin{equation}
 q=\text{Re}(E^\ast j)\ .
\label{qinf}
\end{equation}
The constants $E,j\in \mathbf{C}$ correspond to the complex electric field
$\mathcal{E}$ and current $J$ in
the boundary theory as
\begin{equation}
\mathcal{E}=\left(\frac{\lambda}{2\pi^2}\right)^{1/2}e^{i\Omega t}E
\ ,\qquad
J  = \frac{N_c\sqrt{\lambda}}{2^{5/2}\pi^3}e^{i\Omega t}j
\ ,
\end{equation}
where the real and imaginary parts of $\mathcal{E}$ and $J$ correspond to
$x$- and $y$-components of the electric field and current.
The phase $\theta$ in Eq.~(\ref{horreg}) is closely connected with the
phase of $E$.
If one wants to align the electric field with $x$-axis at $t=0$ as in
Eq.~(\ref{rotEbc}), one can tune the phase $\theta$ so that $E$ becomes a real value. (See section~\ref{BGnum} for details.)

\begin{figure*}[htb]
\centering 
\includegraphics[width=10cm]{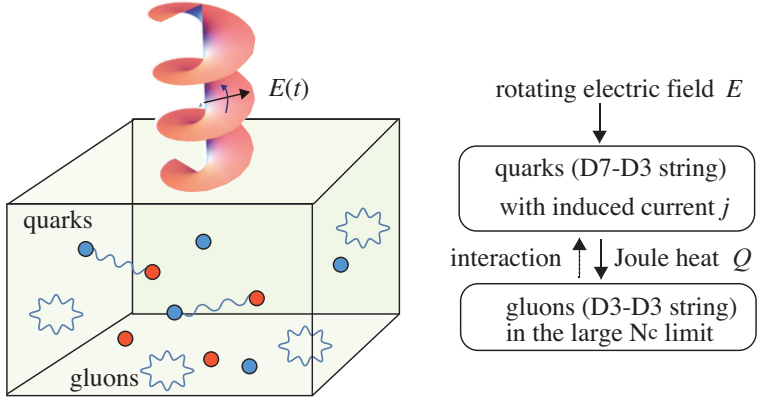}
\caption{
Schematic picture of the energy flow between the degrees of freedom in the system (supersymmetric partners are omitted) : 
The external rotating electric field couples to the quarks and
 anti-quarks, and induces a current. 
 The gluons mediates a long-range 
 interaction between the quarks while they 
act as a heat bath stabilizing the NESS. 
}
\label{fig:schematic}
\end{figure*}

It turns out that the Joule heating is given by 
\begin{equation}
 Q=\textrm{Re}(\mathcal{E}^\ast J)
=\frac{N_c\lambda}{8\pi^4}\textrm{Re}(E^\ast j)
=\frac{N_c\lambda}{8\pi^4} q\ .
\end{equation}
Therefore, the conserved charge $q$ is proportional to the Joule
heating. 
It is worth noting that the conserved current we have seen is nothing
but a steady energy-flux on the D7-brane in the bulk theory.
The rotating electric fields in the boundary theory correspond to the
gauge field similar to a usual circular polarized electromagnetic wave in
the bulk theory (attend to ($t,x,y,\rho$)-components).
Hence, the steady energy-flux like the Poynting flux of the 
circular polarized electromagnetic wave flows from the AdS boundary into
the effective horizon on the D7-brane.
The energy injection from the AdS boundary corresponds to the electric power in the
boundary theory and the ejection to the effective horizon corresponds to
dissipation via the Joule heating.
The flow of energy in this system is schematically summarized in Fig.~\ref{fig:schematic}. 

\subsection{Effective metric and temperature}
\label{sec:Effmet}

Here, we study an effective metric on the D7-brane~\cite{Kim:2011qh,Seiberg:1999vs,Gibbons:2000xe,Gibbons:2001ck,Gibbons:2002tv,Hashimoto:2014yza}.
The effective metric is defined as
\begin{equation}
 \gamma_{ab}=h_{ab}+(2\pi\alpha')^2F_{ac}F_{bd}h^{cd}\ .
\end{equation}
Fluctuations on the brane ``feel'' this effective metric.
For example, if one has found the event horizon with respect to $\gamma_{ab}$,
we cannot probe inside the horizon using the field on the brane.
From Eq.~(\ref{indmet}), (\ref{arealdef}), (\ref{coma}) and (\ref{bdef}), 
the effective metric is written as
\begin{multline}
 \gamma_{ab}d\sigma^a d\sigma^b
=
-F(\rho)dt^2
+\frac{2\Omega\, \textrm{Im}(b'b^\ast)}{\rho^2}dtd\rho
+\frac{1+|b'|^2}{\rho^2}d\rho^2\\
+\frac{\rho^2}{4}(b'e_-+b'{}^\ast e_+
)^2
+\frac{\Omega^2}{4\rho^2}
(b e_-+b^\ast e_+)^2\\
+F(\rho)(e_1^2+e_2^2)
+\rho^2 e_3^2
+d\Omega_3^2\ ,
\label{effmet}
\end{multline}
where 
\begin{equation}
 F(\rho)=\frac{\rho^4-\Omega^2 |b|^2}{\rho^2}\ .
\label{Fdef}
\end{equation}
We have also defined 1-forms $(e_1,e_2,e_3)$ and $e_\pm$ as
\begin{equation}
 \begin{pmatrix}
  e_1 \\
  e_2 \\
  e_3 
 \end{pmatrix}
=
 \begin{pmatrix}
  \cos\Omega t & \sin \Omega t & 0 \\
  -\sin\Omega t & \cos \Omega t & 0 \\
  0 & 0 & 1 
 \end{pmatrix}
 \begin{pmatrix}
  dx \\
  dy \\
  dz 
 \end{pmatrix}
\ ,\qquad e_\pm = e_1\pm ie_2\ .
\end{equation}
Note that $F(\rho)$ becomes zero at $\rho=\rho_c$. 
This implies that $\rho=\rho_c$ is the event horizon with respect to the effective metric $\gamma_{ab}$.%
\footnote{
Let us consider past-directed null geodesics starting from the AdS
boundary for the effective metric $\gamma_{ab}$.
Any tangent vector of the null geodesics must satisfy  
$\gamma_{ab}\frac{d\sigma^a}{ds}\frac{d\sigma^b}{ds} = 0$, where $s$ is
an affine parameter.
Since, from the explicit form of the effective metric (\ref{effmet}),
the norm becomes a sum of squares in terms of components of the tangent
vector other than $(t,\rho)$-directions, we have 
\begin{equation*}
-F(\rho)\left(\frac{dt}{ds}\right)^2
+\frac{2\Omega \textrm{Im}(b'b^\ast)}{\rho^2}\frac{dt}{ds}\frac{d\rho}{ds}
+\frac{1+|b'|^2}{\rho^2}\left(\frac{d\rho}{ds}\right)^2 \le 0 .
\end{equation*}
This means that arbitrary null geodesics on the whole brane worldvolume can be projected to
timelike curves on the $(t,\rho)$-subspace, or they become null only if the
tangent vector of the null geodesics has just the $(t,\rho)$-components. 
Considering the $(t,\rho)$-subspace alone, the locus $\rho=\rho_c$ where
$F(\rho)=0$ is the event horizon, 
namely any past-directed timelike or null curves on the $(t,\rho)$-subspace
cannot enter $\rho\le\rho_c$.
As a result, the past-directed null geodesics on the whole brane
worldvolume cannot enter the region $\rho\le\rho_c$ and the surface
$\rho=\rho_c$ is the event horizon with respect to the effective metric
$\gamma_{ab}$.
}

It turns out that while the coefficients
in the effective metric (\ref{effmet}) are independent of the
time-coordinate $t$ and depend only on $\rho$, the basis
$1$-forms $e_1$ and $e_2$ explicitly depend on $t$.
This means that the effective metric is time-dependent and $\partial_t$ is not a
stationary Killing vector field with respect to $\gamma_{ab}$.
However, on this metric there remains a time-like Killing vector field defined by 
$\chi^a \partial_a = \partial_t + \Omega (x\partial_y - y\partial_x)$.
Indeed, since we can explicitly confirm that the basis $1$-forms satisfy 
$\mathcal{L}_\chi e_1 = 0$ and 
$\mathcal{L}_\chi e_2 = 0$, the effective metric $\gamma_{ab}$ also
satisfies $\mathcal{L}_\chi \gamma_{ab} = 0$.
This Killing vector is a generator of symmetry for other fields such as
the worldvolume gauge field and the induced metric as well as the
effective metric, that is, $\mathcal{L}_\chi a_a = 0$ and
$\mathcal{L}_\chi h_{ab} = 0$.
The conserved current previously shown in (\ref{qjule}) is nothing but a
consequence of this symmetry.
The norm of the Killing vector with respect to the effective metric
becomes
\begin{equation}
 \begin{split}
  \gamma_{ab}\chi^a \chi^b = - F(\rho)
  + \frac{\Omega^2(x^2+y^2)}{2\rho^2}
  (2\rho^2F(\rho) + \rho^4|b'|^2 + \Omega^2|b|^2) \\
  - \frac{\Omega^2}{2\rho^2} 
  \mathrm{Re}[e^{2i\Omega t} (x-iy)^2(\Omega^2 b^2 +\rho^4 b'^2)] .
 \end{split}
\end{equation}
On the surface $\rho=\rho_c$, 
$\chi^a$ is not null vector except for $x=0$ and $y=0$.
The surface $\rho=\rho_c$ is not the Killing horizon generated by the
Killing vector $\chi$.
However, because the system has translational invariance along $x$- and
$y$-directions, even at any $x=x_0$ and $y=y_0$ there exist other 
Killing vectors
which become null at $\rho=\rho_c$ such as 
$\tilde\chi^a \partial_a = \partial_t + \Omega [(x-x_0)\partial_y - (y-y_0)\partial_x]$. 
Although null Killing vectors exist everywhere on the event horizon,
these Killing vectors do not belong to a single Killing vector field.

The effective metric is explicitly regular at $\rho=\rho_c$.
The Hawking temperature is computed from the surface gravity $\kappa$ on
the event horizon as
\begin{equation}
 T_H=\frac{\kappa}{2\pi}=
\frac{2\rho_c-\Omega \, \textrm{Re}\, p}{2\pi \, \textrm{Im}\, p}\ ,
  \label{TH}
\end{equation}
where $p$ is given in Eq.~(\ref{horreg}).
Here, the surface gravity we have defined above is different from the usual
definition in the sense that the event horizon of the effective geometry
is not a Killing horizon.
However, since the null Killing vectors associated with the
event horizon exist as we mentioned, we can define a surface gravity on
the event horizon with respect to these Killing vectors.
For example, we have $\chi^aD_a\chi^b = \kappa \chi^b$ on the event
horizon for a Killing vector $\chi^a$ satisfying 
$\chi^a\chi^b \gamma_{ab}|_{\rho=\rho_c}=0$.
Moreover, if we restrict our attention to the $(t,\rho)$-subspace, 
the event horizon $\rho=\rho_c$ can be regarded as the Killing horizon
on the subspace because the subspace is time-independent.
We can also obtain the above surface gravity $\kappa$ in the same way as the
surface gravity with the Killing horizon for the $(t,\rho)$-subspace.
Quantum fluctuations which can reach the AdS boundary from neighborhood
of the event horizon are dominated by the $s$-wave modes.
In other words, null geodesics with no components other than $(t,\rho)$
can mainly arrive at the boundary.
The Hawking
temperature of the current system can be characterized by this surface gravity.

In Eq.~(\ref{qofx}), the Joule heating $q$ is expressed by the function
of $\rho_c$ as $q=\rho_c^5$. Thus, replacing $\rho_c$ by $q^{1/5}$ in Eq.~(\ref{TH}),
we can express the Joule heating as a function of $T_H$.
For $T_H\gg \Omega$ and $T_H \ll \Omega$,
the Joule heating $q$ is simply written as
\begin{equation}
 q=\left(\frac{4 \pi}{5}T_H\right)^5\ ,\quad(T_H \ll \Omega)\ ,
\qquad
 q=\left(\frac{2\pi}{\sqrt{6}} T_H\right)^5\ ,\quad(T_H \gg \Omega)\ .
\end{equation}

We introduce coordinates $(\tau, \rho_\ast)$ as
\begin{equation}
d\tau=dt-\frac{\Omega\, \textrm{Im}(b'b^\ast)}{\rho^2 F(\rho)}d\rho\
  ,\quad
  d\rho_\ast=\frac{\mathcal{L}_0}{\rho^3 F(\rho)} d\rho\ .
  \label{tau_rhoast}
\end{equation}
The range of the tortoise coordinate $\rho_\ast$ is $-\infty < \rho_\ast \leq 0$.
In term of these coordinates, $(t,\rho)$-part of the effective metric
can be  
diagonalized as
\begin{multline}
 \gamma_{ab}d\sigma^a d\sigma^b
=
F(\rho)(-d\tau^2+d\rho_\ast^2)
+\frac{\rho^2}{4}(b'e_-+b'{}^\ast e_+
)^2
+\frac{\Omega^2}{4\rho^2}
(b e_-+b^\ast e_+)^2\\
+F(\rho)(e_1^2+e_2^2)
+\rho^2 e_3^2
+d\Omega_3^2\ .
\end{multline}
In section~\ref{sec:Hall}, 
we will find that the new coordinate $(\tau,\rho_\ast)$ are better suited to describe the perturbation of the background solution $b(\rho)$.

\subsection{Physical quantities of the holographic Floquet state}
\label{BGnum}

\begin{figure}
  \centering
  \subfigure[Varying $|E|$ for fixed $\Omega$]
  {\includegraphics[scale=0.5]{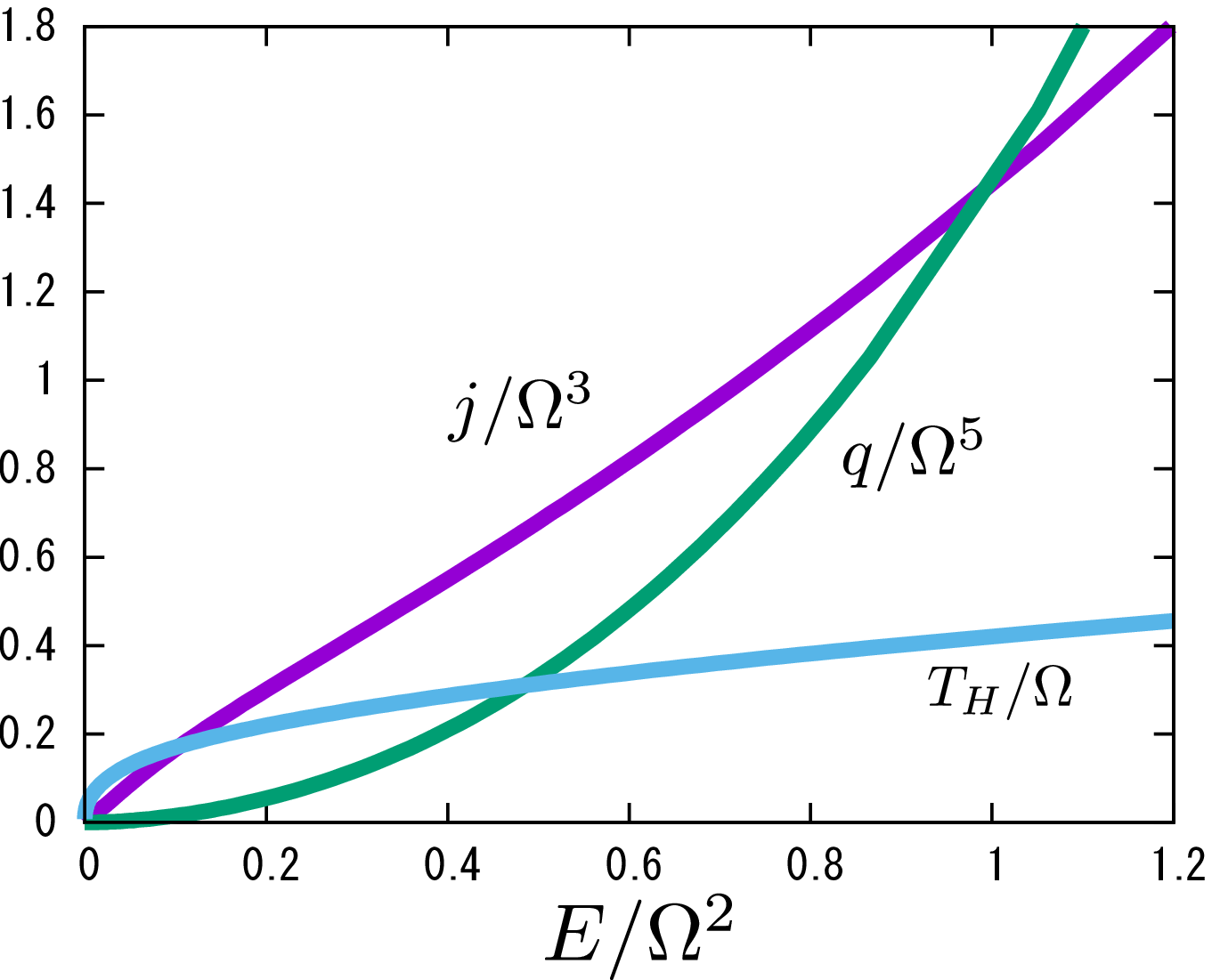}\label{JQE}
   }
  \subfigure[Varying $\Omega$ for fixed $E$]
  {\includegraphics[scale=0.5]{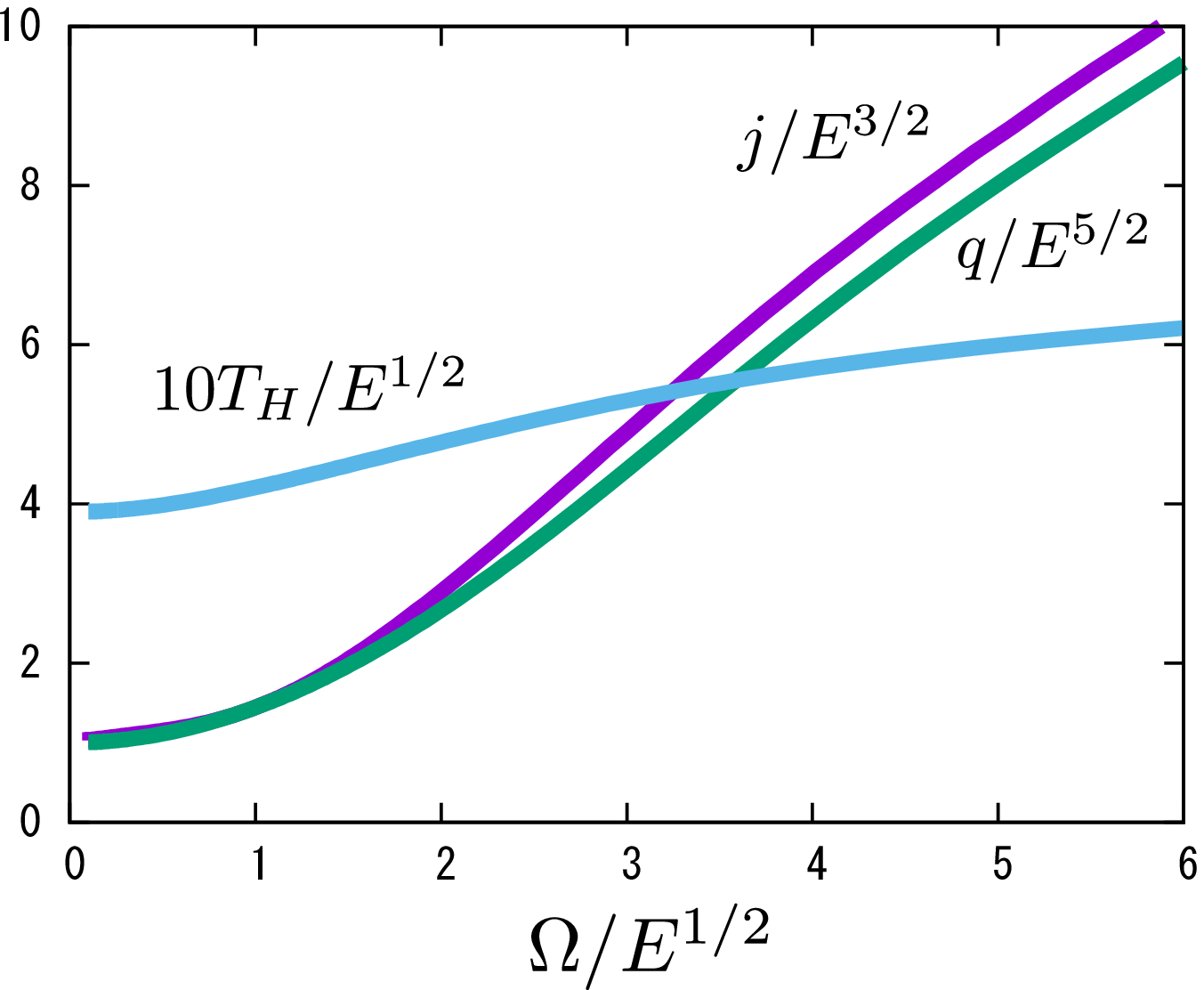}\label{JQOm}
  }
  \caption{
 Electric current, Joule heating and Hawking temperature.
 They are increasing functions of $E$ and $\Omega$ for fixed $\Omega$ and $E$, respectively.
 In the right figure, we multiply $10$ to Hawking temperature $T_H$
 for visibility.
}
\label{JQE00}
\end{figure}

Here, we determine the $b(\rho)$ numerically and evaluate the physical quantities in the boundary theory.
Note that 
our system is invariant under the scaling symmetry:
\begin{equation}
\begin{split}
 &\rho\to \lambda \rho,\quad
 t\to t/\lambda,
 \quad
 \Omega\to \lambda\Omega,
 \quad
 b\to \lambda b,\\
 &E\to \lambda^2 E,\quad j\to \lambda^3 j,\quad
 q\to \lambda^5 q,\quad
 T_H \to \lambda T_H,
\end{split}
\end{equation}
where $\lambda$ is a non-zero constant.
Using the scaling symmetry, we will set $\Omega=1$ in our numerical calculation. 
We solve Eq.~(\ref{beq}) from $\rho=\rho_c+\delta$ to $\rho=\rho_\textrm{max}$, where
we typically set $\delta=1.0\times 10^{-6}$ and $\rho_\textrm{max}=100$.
At the inner boundary $\rho=\rho_c+\delta$,
we impose the boundary condition in Eq.~(\ref{horreg}). 
At first, we set $\theta=0$ tentatively and obtain a solution
$b_0(\rho)$.
From numerical values of $b_0(\rho_\textrm{max})$ and $b'_0(\rho_\textrm{max})$,
we read off $E_0, j_0\in \mathbf{C}$ using Eq.~(\ref{binf}).
To make $E$ real, we choose 
$\theta=-\arg(E_0)$ and obtain a desirable solution
$b(\rho)=e^{i\theta}b_0(\rho)$, which satisfies the boundary
condition~(\ref{rotEbc}).
We repeat above procedure for several $\rho_c$ and obtain $E(\rho_c)\in \mathbf{R}$ and $j(\rho_c)\in \mathbf{C}$.
We can also compute the Joule heating $q$ and Hawking temperature $T_H$ as functions of $\rho_c$
from Eqs.~(\ref{qinf}) and (\ref{TH}).
In Fig.~\ref{JQE}, we show scaling invariant quantities $|j|/\Omega^3$, $q/\Omega^5$ and $T_H/\Omega$ as
functions of $E/\Omega^2$ regarding $\rho_c$ as a parameter.
On the other hand, in Fig.~\ref{JQOm}, we show the same results in different normalization: 
$|j|/E^{3/2}$, $q/E^{5/2}$ and $T_H/E^{1/2}$ vs $\Omega/E^{1/2}$.
Although these figures are essentially same, they are convenient to see two kinds of physical process: 
{\it{varying $E$ for fixed $\Omega$}} and {\it{varying $\Omega$ for fixed $E$}}.
In the left figure, we see that
$|j|$, $q$ and $T_H$ increase as functions of $E$ for fixed $\Omega$ as one can expect.
In the right figure, however, we find non-trivial feature: 
$|j|$, $q$ and $T_H$ are also increasing functions of the frequency
$\Omega$ for fixed $E$.
It follows that, the more quickly we rotate the electric field,
the more effectively the electric current is generated.
The vertical axis $\Omega=0$ in the right figure corresponds to
the direct current (DC) electric field, or more precisely a static
electric field.
The D3/D7 systems with the DC electric fields 
were studied in Refs.~\cite{Karch:2007pd,Albash:2007bq,Erdmenger:2007bn,Hashimoto:2013mua,Hashimoto:2014yza}.
In our notation, the electric current has been given by $j=E^{3/2}$, $q=E^{5/2}$ and
$T_H=(6E)^{1/2}/(2\pi)\simeq 0.39E^{1/2}$. 
Therefore, our results for the rotating electric fields can consistently
reproduce results for the static electric fields in the limit
$\Omega \to 0$.

\section{Hall effect of the holographic Floquet state}
\label{sec:Hall}

\subsection{Conductivities}
\label{sec:sigmas}

In this section, we study the linear response of the
holographic Floquet state 
against 
probe AC and DC electric fields.
We apply the probe AC electric field as\footnote{
For perturbation theory, we will use the vector notation instead of the complex notation.
}
\begin{equation}
 \vec{\varepsilon}(t)=\vec{\varepsilon}_\omega e^{-i\omega t}\ .
\end{equation}
By the probe electric field, the gauge field on the brane is perturbed as 
$\vec{a}\to \vec{a} + \delta \vec{a}$.
The boundary condition for the perturbation of gauge field $\delta \vec{a}$ becomes
\begin{equation}
 \delta \vec{a}|_{\rho=\infty} = -\frac{i \vec{\varepsilon}_\omega}{\omega} 
e^{-i\omega  t}\ .
\label{delta_a_b}
\end{equation}
For the perturbation of $\vec{b}$,
the boundary condition is written as 
\begin{equation}
\delta \vec{b}|_{\rho=\infty} =
\begin{pmatrix}
  \cos\Omega t & \sin \Omega t \\
  -\sin\Omega t & \cos \Omega t 
 \end{pmatrix}
\delta \vec{a}|_{\rho=\infty}
=-\frac{i}{\omega}(\bm{M}_- e^{-i\omega_- t}+\bm{M}_+ e^{-i\omega_+
t})\vec{\varepsilon}_\omega \ .
\label{deltab_b}
\end{equation}
where $\omega_\pm=\omega\pm \Omega$ and $\bm{M}_\pm$ is a constant matrix
defined by
\begin{equation}
 \bm{M}_\pm=
\frac{1}{2}
\begin{pmatrix}
  1 & \pm i \\
  \mp i & 1 
\end{pmatrix}
\ .
\end{equation}
The perturbation equation for $\delta\vec{b}$ is written as
\begin{equation}
 [\partial_\tau^2-\partial_{\rho_\ast}^2
+\bm{A}(\rho)\partial_\tau
+\bm{B}(\rho)\partial_{\rho_\ast}
+\bm{C}(\rho)]\delta \vec{b}=0\ ,
\label{vecbeq}
\end{equation}
where we have used coordinates $(\tau, \rho_\ast)$ defined in Eq.~(\ref{tau_rhoast}) and 
$\bm{A}$, $\bm{B}$ and $\bm{C}$ are $\rho$-dependent $2\times 2$ matrices.
Their explicit expressions are summarized in appendix~\ref{app:pert}.
From Eq.~(\ref{deltab_b}), 
fluctuations with frequencies $\omega_+$ and $\omega_-$ are induced by
the boundary conditions. Thus,
$\delta \vec{b}$ is expanded as
\begin{equation}
 \delta\vec{b}= \vec{\beta}_+(\rho) e^{-i\omega_+ t} +
  \vec{\beta}_-(\rho) e^{-i\omega_- t}\ .
\end{equation}
From Eq.~(\ref{vecbeq}), we obtain decoupled equations for 
$\vec{\beta}_\pm$ as
\begin{equation}
\left[
\frac{d^2}{d\rho_\ast^2} 
-\bm{B}(\rho)\frac{d}{d\rho_\ast} 
+\omega_\pm^2
+i\omega_\pm \bm{A}(\rho) 
-\bm{C}(\rho)\right]\vec{\beta}_\pm=0\ .
\label{beta_eq}
\end{equation}
Near the AdS boundary $\rho_\ast=0$, asymptotic forms of $\vec{\beta}_\pm$ are 
\begin{equation}
 \vec{\beta}_\pm(\rho)
= \vec{\beta}^{\,(0)}_\pm 
+ \vec{\beta}^{\,(2)}_\pm \, \rho_\ast^2 
- \frac{1}{2}(\omega_\pm^2+\Omega^2-2i\Omega\omega_\pm \bm{\epsilon})\vec{\beta}^{\,(0)}_\pm \, \rho_\ast^2
\ln(-\Omega \rho_\ast)+\cdots\ ,
\label{beta_inf}
\end{equation}
where 
$\vec{\beta}^{\,(0)}_\pm$ and $\vec{\beta}^{\,(2)}_\pm$ are constant
vectors and 
$\bm{\epsilon}$ is the anti-symmetric matrix with $\bm{\epsilon}_{12}=1$.
From the boundary condition~Eq.~(\ref{deltab_b}), the leading term must be
\begin{equation}
 \vec{\beta}^{\,(0)}_\pm =-\frac{i}{\omega}\bm{M}_\pm \vec{\varepsilon}_\omega \ ,
\label{betainf}
\end{equation}
On the other hand, we impose the ingoing wave boundary condition at the
horizon:
\begin{equation}
 \vec{\beta}_\pm\propto e^{-i\omega_\pm \rho_\ast}\ ,\quad
 (\rho\to\rho_c)\ .
\label{betahor}
\end{equation}
The asymptotic solution of $\delta \vec{b}$ near the horizon is
studied in appendix~\ref{app:pert}.
From Eq.~(\ref{beta_inf}), we can also write down the asymptotic solution of
the original gauge field $\delta \vec{a}$ near the AdS boundary.
Using Eq.~(\ref{ainf}), we can read off the electric current induced by the probe electric field as
\begin{equation}
 \delta\vec{j}_\omega=2e^{-i\omega t}\left[(\bm{M}_+ \vec{\beta}_+^{\,(2)}+\bm{M}_- \vec{\beta}_-^{\,(2)})
+\bm{M}_+ \vec{\beta}_-^{\,(2)} e^{2i\Omega t}
+\bm{M}_- \vec{\beta}_+^{\,(2)} e^{-2i\Omega t}\right]\ .
\label{j_beta2}
\end{equation}
Even though we have introduced the monochromatic AC electric field with the frequency $\omega$
as in Eq.~(\ref{delta_a_b}),
three modes are induced in the electric current, whose frequencies are
given by $\omega$ and $\omega\pm 2\Omega$.
This is known as the heterodyning effect characteristic to periodically driven Floquet systems \cite{Oka2016}.
Thus, we can define three kinds of conductivities $\bm{\sigma}$ and
$\bm{\sigma}^\pm$ as\footnote{
Again there is an ambiguity in the electric current
$\delta\vec{j}\to \delta \vec{j}+\alpha \dot{\vec{\varepsilon}}$, where $\alpha$ is a constant.
This ambiguity appears in the conductivity as
$\bm{\sigma}\to \bm{\sigma}-i\alpha \omega$. Thus, only 
$\textrm{Im}\,\sigma_{xx}$ and $\textrm{Im}\,\sigma_{yy}$ are affected by the ambiguity in the conductivities.
}
\begin{equation}
 \delta \vec{j}_\omega=\left[\bm{\sigma}(\omega)e^{-i\omega t} 
+\bm{\sigma}^+(\omega)e^{^{-i(\omega+2\Omega) t}}+\bm{\sigma}^-(\omega)e^{-i(\omega-2\Omega)t}\right]
 \vec{\varepsilon}_\omega \ ,
\label{sigma_def}
\end{equation}
where $\bm{\sigma}$'s are $2\times 2$ complex matrices.
Conductivities $\bm{\sigma}$ and $\bm{\sigma}^\pm$ have 
$4\times 3=12$ complex components. However, they are not independent but
there are the following 8 relations:
\begin{equation}
\begin{split}
&\sigma_{xx}=\sigma_{yy}\ ,\qquad \sigma_{xy}=-\sigma_{yx}\ ,\\
&\sigma^\pm_{xx}=-\sigma^\pm_{yy}\ ,\qquad
 \sigma^\pm_{xy}=\sigma^\pm_{yx}\ ,\qquad
\sigma^\pm_{yx}=\pm i \sigma^\pm_{xx}\ .
\end{split}
\label{sigmasym}
\end{equation}
We give a proof of these relations in appendix~\ref{sym_cund}.
Therefore, independent complex degrees of freedom of 
conductivities are $12-8=4$. 
We will take $\sigma_{xx}$, $\sigma_{xy}$ and $\sigma^\pm_{xx}$ as
the independent components.
Taking the complex conjugate of Eq.~(\ref{sigma_def}) and using reality conditions
$\delta j_{-\omega}=\delta j_\omega^\ast$ and $\varepsilon_{-\omega}=\varepsilon_\omega^\ast$, we obtain
\begin{equation}
\bm{\sigma}(-\omega)=(\bm{\sigma}(\omega))^\ast\ ,\quad
 \bm{\sigma}^\pm(-\omega)=(\bm{\sigma}^\mp(\omega))^\ast\ .
 \label{sigma_cc}
\end{equation}
So, we will consider the conductivities only in $\omega\geq 0$.

\subsection{DC Hall effect}
\label{sec:DCHall}

Here, we study the DC conductivity in the holographic Floquet state obtained in section~\ref{sec:FWS}.
In the previous subsection,
we have considered the perturbation of the gauge field in the frequency domain.
However,
it is not directly applicable for the probe DC electric field since Eq.~(\ref{delta_a_b}) becomes singular at $\omega=0$.
Therefore, we take the time-domain approach to evaluate the DC conductivities.
We will find that the DC conductivities computed in the time domain 
will coincide with the AC conductivities computed in the frequency
domain when taking the limit of $\omega\to 0$.

\begin{figure}
  \centering
  \subfigure[Varying $E$ for fixed $\Omega$]
  {\includegraphics[scale=0.45]{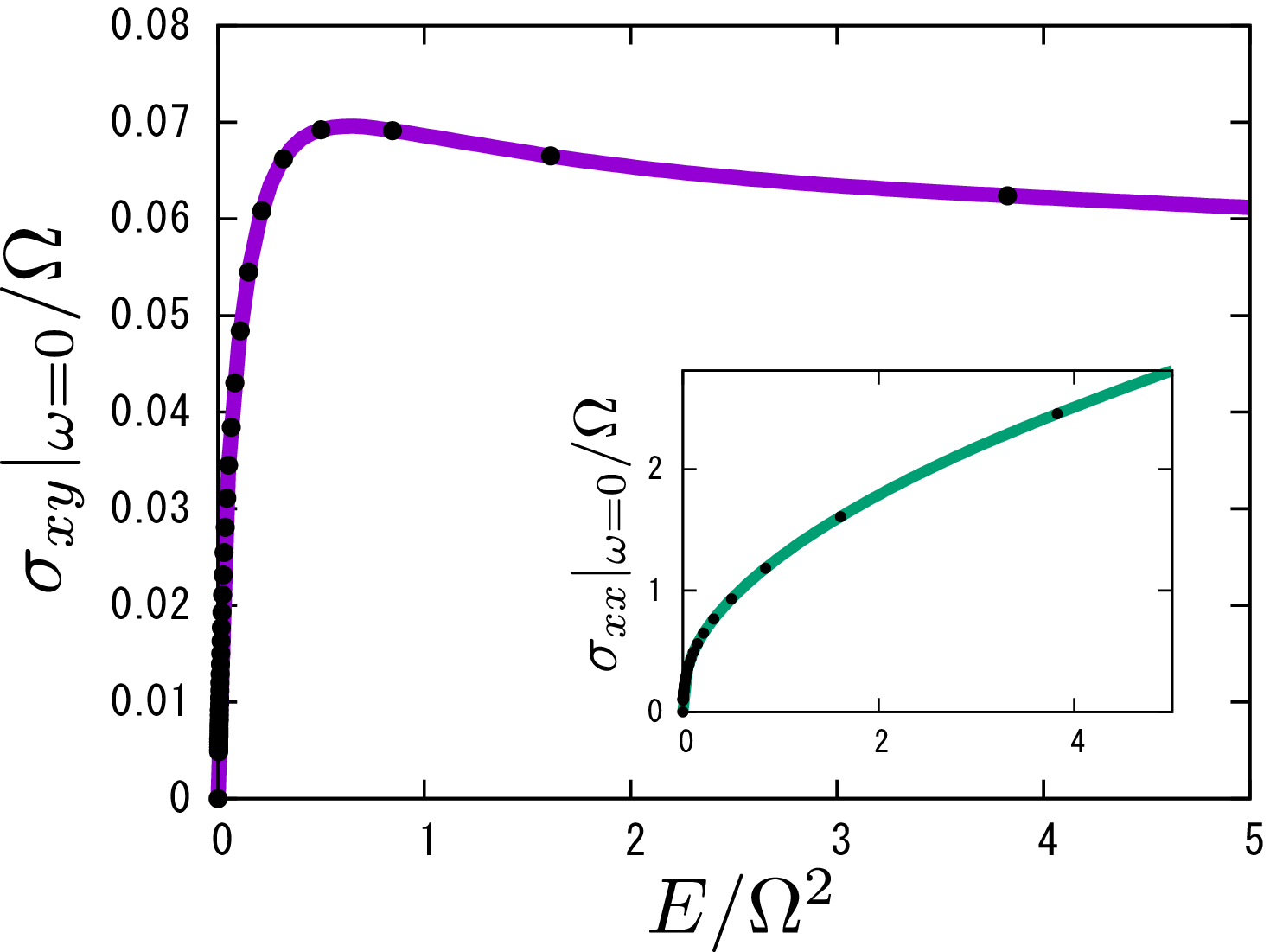}\label{syxE}
   }
  \subfigure[Varying $\Omega$ for fixed $E$]
  {\includegraphics[scale=0.45]{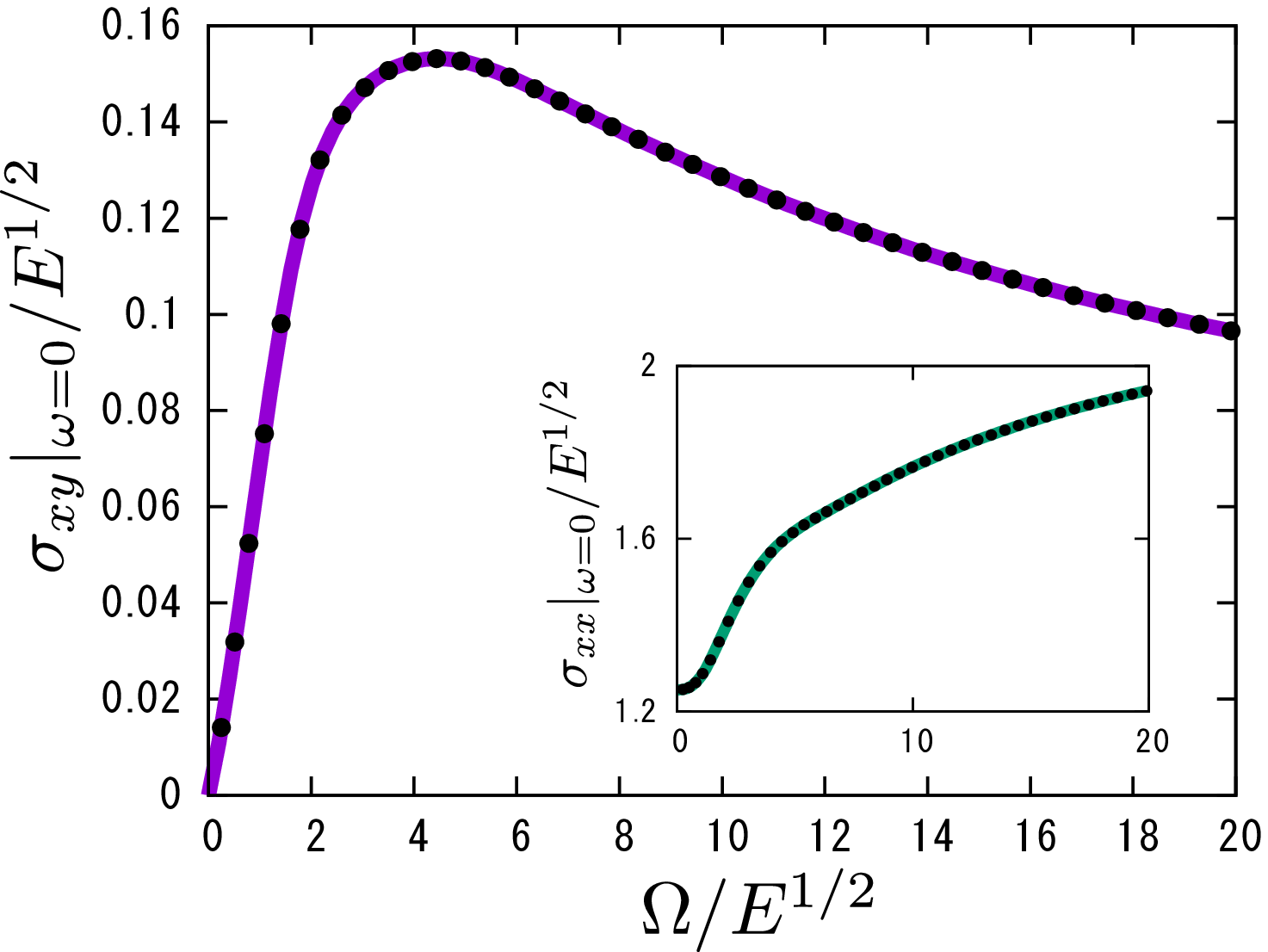}\label{syxOm}
  }
  \caption{
The DC Hall conductivity of the holographic Floquet state.
In the left and right figures, parameter are nondimensionalized by $\Omega$ and $E$, respectively.
They have maximum values at 
 $(E/\Omega^2,\sigma_{xy}|_{\omega=0}/\Omega)=(0.639,0.0696)$ [left] and
 $(\Omega/E^{1/2},\sigma_{xy}|_{\omega=0}/E^{1/2})=(4.39,0.153)$ [right].
 The insets of these figures are for $\sigma_{xx}$. 
}
\label{syx}
\end{figure}

We consider a quench-type function for the probe electric field as
\begin{equation}
\vec{\varepsilon}(t)= \vec{\varepsilon}_f f(t)\ ,\quad
f(t)\equiv 
\begin{cases}
0 & (t<0) \\
[t-\frac{\Delta t}{2\pi}\sin(2\pi t/\Delta t)]/\Delta t & 
 (0 \le t \le \Delta t) \\
1 & (t>\Delta t) 
\end{cases}
\ .
\label{Efunc}
\end{equation}
Here, $\vec{\varepsilon}_f$ is a final value of the electric field and
$\Delta t$ is a duration of the quench.
The boundary condition at the AdS boundary for $\delta \vec{b}$ becomes
\begin{equation}
 \delta \vec{b}|_{\rho=\infty}=-
  \begin{pmatrix}
  \cos\Omega t & \sin \Omega t \\
  -\sin\Omega t & \cos \Omega t 
   \end{pmatrix}\int^t_0 dt' \vec{\varepsilon}(t')\ .
\label{dbbdry}
\end{equation}
The numerical method to solve the perturbation equations in the time domain is summarized in appendix~\ref{app:num}.
From the numerical solution of $\delta \vec{b}(t,\rho)$, we can compute the
perturbation of the original gauge field, $\delta \vec{a}(t,\rho)$.
The asymptotic form of $\delta \vec{a}(t,\rho)$ near the AdS boundary is written as
\begin{equation}
 \delta \vec{a}(t,\rho)=-\int^t_0 dt' \vec{\varepsilon}(t')
+\frac{\delta \vec{j}(t)}{2\rho^2}+\frac{\dot{\vec{\varepsilon}}(t)}{2\rho^2}\ln\left(\frac{\rho}{\Omega}\right)
\ ,
\end{equation}
where $\delta \vec{j}(t)$ is the perturbation of the electric current.
From the numerical solution, 
we read off the electric current induced by the probe electric field
$\vec{\varepsilon}$.
Taking the limit of $\omega\to 0$ in Eq.~(\ref{sigma_def}), 
we obtain the electric current at late time as
$\delta \vec{j}=[\bm{\sigma}(0)+\bm{\sigma}^+(0)e^{^{-2i\Omega t}}+\bm{\sigma}^-(0)e^{2i\Omega t}]\vec{\varepsilon}_f$.
Note that, from Eq.~(\ref{sigma_cc}),
$\bm{\sigma}(0)$ becomes a real matrix
and $\bm{\sigma}^+(0)$ is the complex conjugate of $\bm{\sigma}^-(0)$.
Combining them with Eq.~(\ref{sigmasym}), 
we can take the independent components of the conductivities as
$\sigma_{xx}(0), \sigma_{xy}(0)\in \mathbf{R}$ and $\sigma_{xx}^+(0)\in \mathbf{C}$ (four real components).
In the actual numerical calculation, we set $\vec{\varepsilon}_f=(1,0)$. Then, the late time expression of the current is written as 
\begin{equation}
 \delta \vec{j}=
 \begin{pmatrix}
 \sigma_{xx}(0) \\
 -\sigma_{xy}(0)
 \end{pmatrix}
 +2
 \begin{pmatrix}
 \textrm{Re}\,\sigma_{xx}^+(0) \\
 -\textrm{Im}\,\sigma_{xx}^+(0) 
 \end{pmatrix}
 \cos 2\Omega t
 +2
 \begin{pmatrix}
 \textrm{Im}\,\sigma_{xx}^+(0) \\
 \textrm{Re}\,\sigma_{xx}^+(0)
 \end{pmatrix}
 \sin 2\Omega t
 \ .
 \label{jlate}
\end{equation}
\begin{figure}
\begin{center}
\includegraphics[scale=0.5]{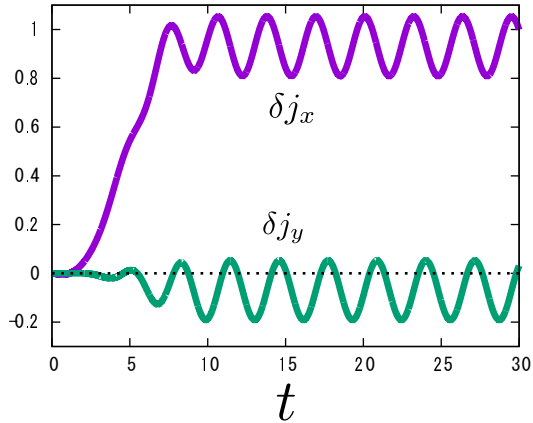}
\end{center}
\caption{
Time dependence of the electric currents for $E=0.498$,  
 $\Omega=1$, $\Delta t=10$ and $\vec{\varepsilon}_f=(1,0)$.
The time-average of the $\delta j_y$ at the late time
has a non-zero value.
}
 \label{dJ}
\end{figure}
In Fig.~\ref{dJ}, we show the time dependence of 
$\delta j_x$ and $\delta j_y$ 
for $E/\Omega^2=0.498$ and $\Omega\Delta t=10$ as an example.
Note that,
even though the probe DC electric field $\vec{\varepsilon}$ does not have oscillating components at late time,
we obtain oscillating response currents because of the background rotating electric field. 
We can find that the time-average of the $\delta j_y$ at the late time
has a non-zero negative value. This is nothing but the Hall effect induced by the rotating
external electric field. The oscillations of $\delta j_x$ and $\delta j_y$ come from $\cos 2\Omega t$ and $\sin 2\Omega t$ in Eq.~(\ref{jlate}).
When we apply the probe DC electric field along $x$-direction, we obtain the Hall current towards $(-y)$-direction. (In other wards, $\sigma_{xy}|_{\omega=0}$ is positive.) This is one of clear predictions to strongly coupled version of ``Floquet Weyl semimetals.''
We fit out the numerical result of $\delta\vec{j}(t)$ using
Eq.~(\ref{jlate}) and determine the conductivities.
In Fig.~\ref{syx}, we show the DC Hall conductivity $\sigma_{xy}|_{\omega=0}$
as the function of background parameters:
(a) $\sigma_{xy}|_{\omega=0}/\Omega$ vs $E/\Omega^2$, 
and (b) $\sigma_{xy}|_{\omega=0}/E^{1/2}$ vs $\Omega/E^{1/2}$.
In the both figures, there are maximum values in the Hall conductivity.
When we vary $E$ for a fixed $\Omega$,
the DC Hall conductivity has a maximum 
value $\sigma_{xy}|_{\omega=0}=0.0696\Omega$ at $E=0.639\Omega^2$. On the other hand, 
when we vary $\Omega$ for a fixed $E$, it has a maximum
value $\sigma_{xy}|_{\omega=0}=0.153E^{1/2}$ at $\Omega=4.39 E^{1/2}$. 
What is the reason that the Hall conductivity peaks out and start to decrease? 
Currently, we can only make speculations. In
section~\ref{sec:weakcoupling}, we have shown that
in the weak coupling coupling limit, the Weyl points, which causes the Hall response, 
vanishes at $E=0.5\Omega^2$. This occurs through a pair annihilation of Weyl points with the 
new Weyl points that emerges from Floquet side bands \cite{Oka-new}. 
The value of the rotating electric field  $E=0.639\Omega^2$ 
where the Hall response peaks out is close to $E=0.5\Omega^2$ 
and thus the  disappearance of the Weyl point may be the cause. 
Another possibility is through destruction of the Weyl band due to interaction. 
For example, in Ref.~\cite{Morimoto:2016}, it was pointed out that a gap may open in a single Weyl point when there are
fermion-fermion interactions.

In the insets of these figures, we also show the $\sigma_{xx}$.
They are increasing functions of $E$ and $\Omega$ for fixed $\Omega$ and $E$, respectively.
In $(2+1)$ dimensions. free systems~\cite{OkaAoki},
however, it was found that $\sigma_{xx}$ increases for small $E$ but decreases for large $E$.
This discrepancy would be originate from the dimensionality.
In $(2+1)$ dimensions, the rotating electric field opens a gap at the Dirac point.
Therefore, for stronger rotating electric fields, the system tends to be more like an insulator.
On the other hand, in $(3+1)$ dimensions,
it does not open the gap at least for small $E$ but just splits the
Dirac point into Weyl points.
Therefore, $\sigma_{xx}$ becomes an increasing function of $E$ due to carriers induced by the rotating electric field.

The other conductivities $\bm{\sigma}^\pm$ that we have also studied are summarized in appendix~\ref{app:other}.

\subsection{Optical Hall effect}
\label{sec:ACHall}

We study the conductivities for AC probe electric fields ($\omega>0$) 
using the frequency-domain approach shown in section~\ref{sec:sigmas}.
Technical details are as follows.
We solve Eq.~(\ref{beta_eq}) from the effective horizon
($\rho_\ast=-\infty$) to the AdS boundary ($\rho_\ast=0$) using the fourth order Runge-Kutta method. At the effective horizon,
we impose the boundary conditions $\vec{\beta}_\pm = \vec{c}_\pm e^{-i\omega_\pm \rho_\ast}$ as in Eq.~(\ref{betahor}) where $\vec{c}_\pm$ are constant vectors. 
We solve the perturbation equation twice for two initial conditions: $\vec{c}_\pm=(1,0)$ and  $(0,1)$.
From asymptotic forms of the numerical solutions, we read off $\vec{\beta}^{\,(0)}_\pm$ and $\vec{\beta}^{\,(2)}_\pm$ defined in Eq.~(\ref{beta_inf}).
Because of the linearity, we obtain complex $2\times 2$ matrices $P_\pm$ and $Q_\pm$ defined by
$\vec{\beta}^{\,(0)}_\pm=P_\pm \vec{c}_\pm$ and $\vec{\beta}^{\,(2)}_\pm=Q_\pm \vec{c}_\pm$.
Then, we can express $\vec{\beta}^{\,(2)}_\pm$ by $\vec{\beta}^{\,(0)}_\pm$ as
$\vec{\beta}^{\,(2)}_\pm=Q_\pm P_\pm^{-1} \vec{\beta}^{\,(0)}_\pm$.
Using Eq.~(\ref{betainf}), we have $\vec{\beta}^{\,(2)}_\pm=-(i/\omega) Q_\pm P_\pm^{-1} \bm{M}_\pm \vec{\varepsilon}_\omega$.
Substituting this into Eq.~(\ref{j_beta2}) and reading off coefficients of $e^{-i\omega t}$ and $e^{-i(\omega\pm2\Omega) t}$,
we obtain the AC conductivities $\bm{\sigma}$ and $\bm{\sigma}^\pm$.

\begin{figure}
  \centering
  \subfigure[Varying $E$ for fixed $\Omega$]
  {\includegraphics[scale=0.45]{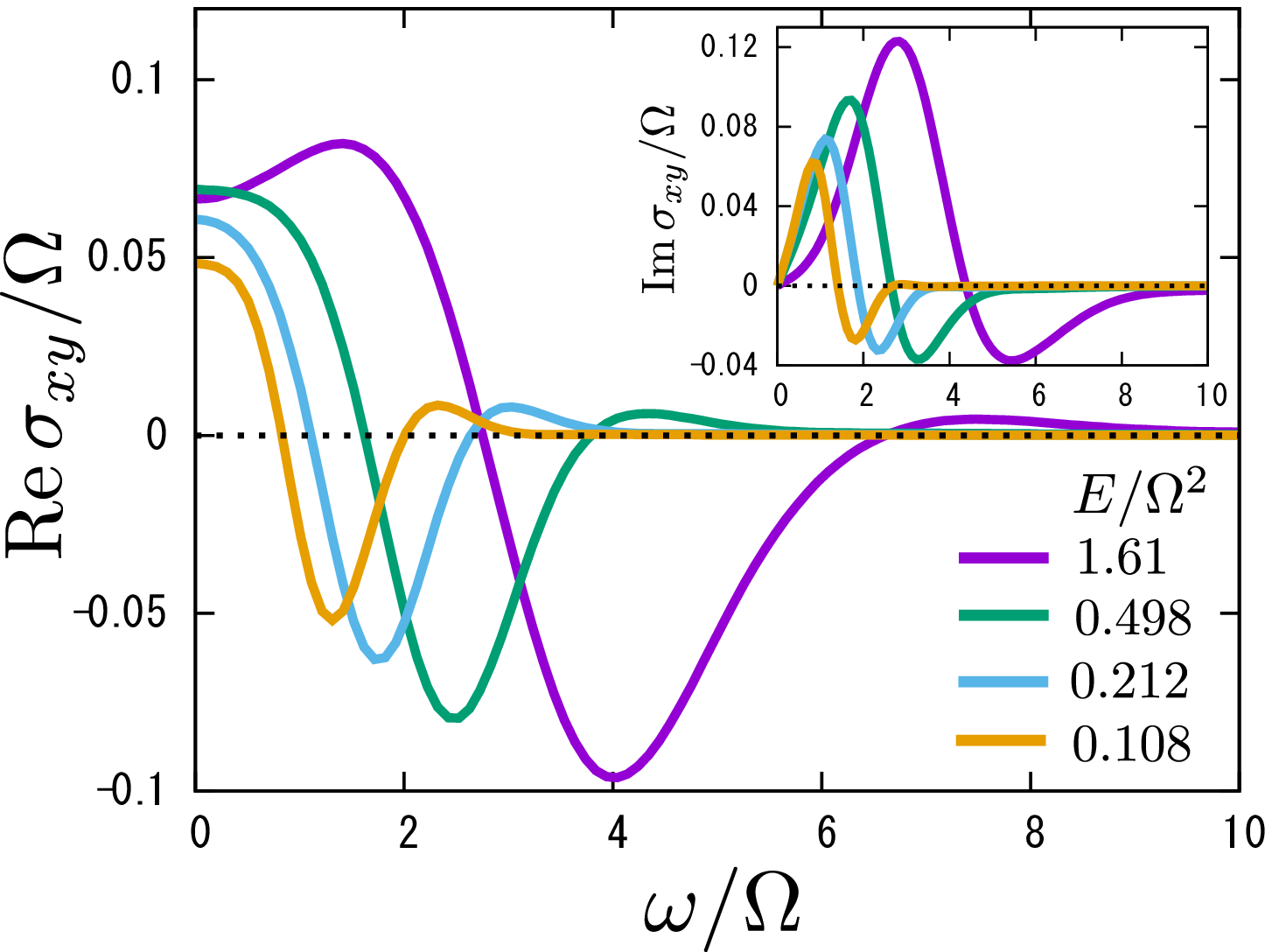}
   }
  \subfigure[Varying $\Omega$ for fixed $E$]
  {\includegraphics[scale=0.45]{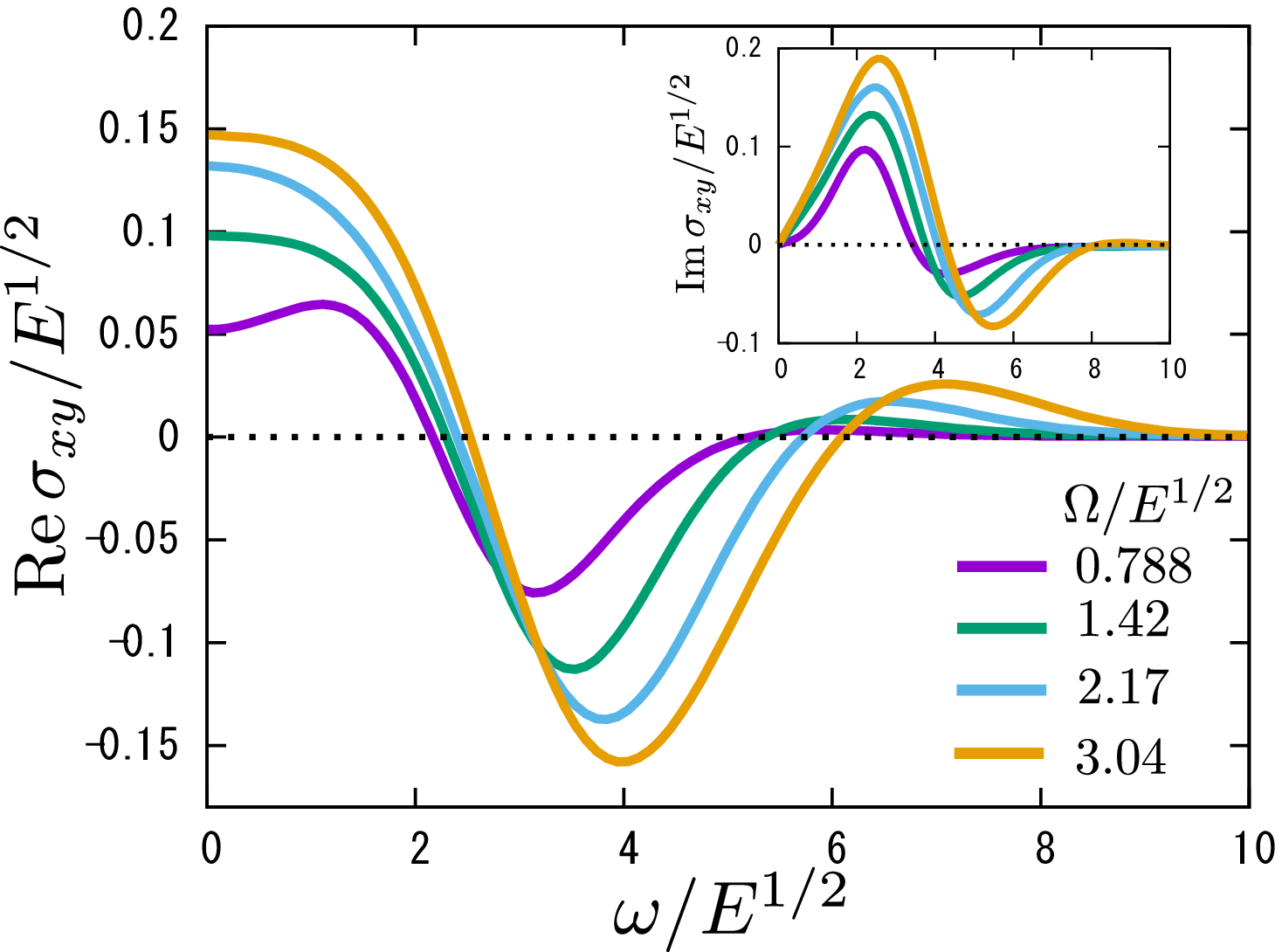}
  }
 \caption{
 The optical Hall coefficients against the probe AC frequency $\omega$.
 Curves in the left (right) figure correspond to several $E$ ($\Omega$)
 for a fixed $\Omega$ ($E$).
}
\label{Sxy_AC}
\end{figure}

\begin{figure}
  \centering
  \subfigure[Varying $E$ for fixed $\Omega$]
  {\includegraphics[scale=0.45]{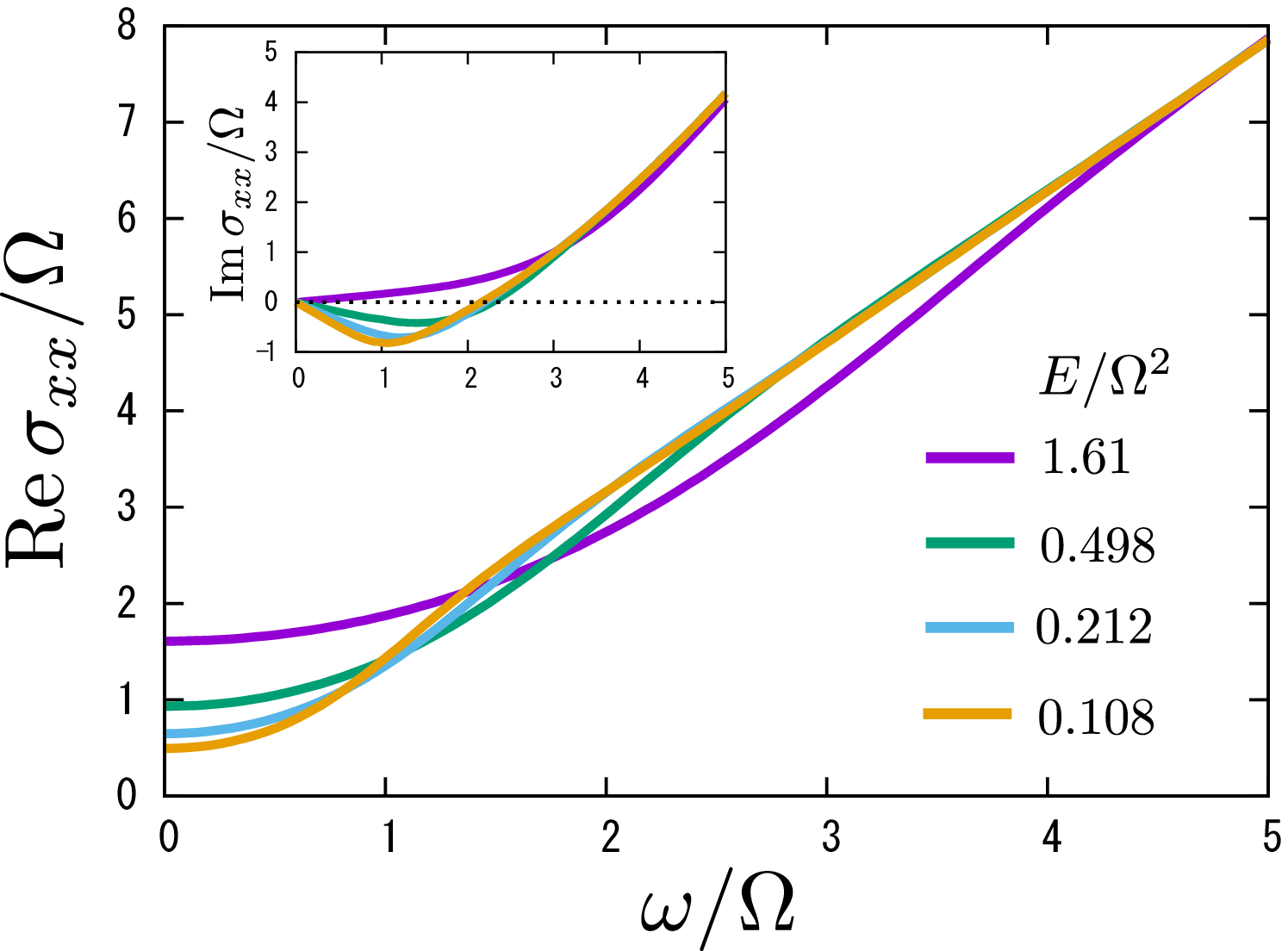}
   }
  \subfigure[Varying $E$ for fixed $\Omega$]
  {\includegraphics[scale=0.45]{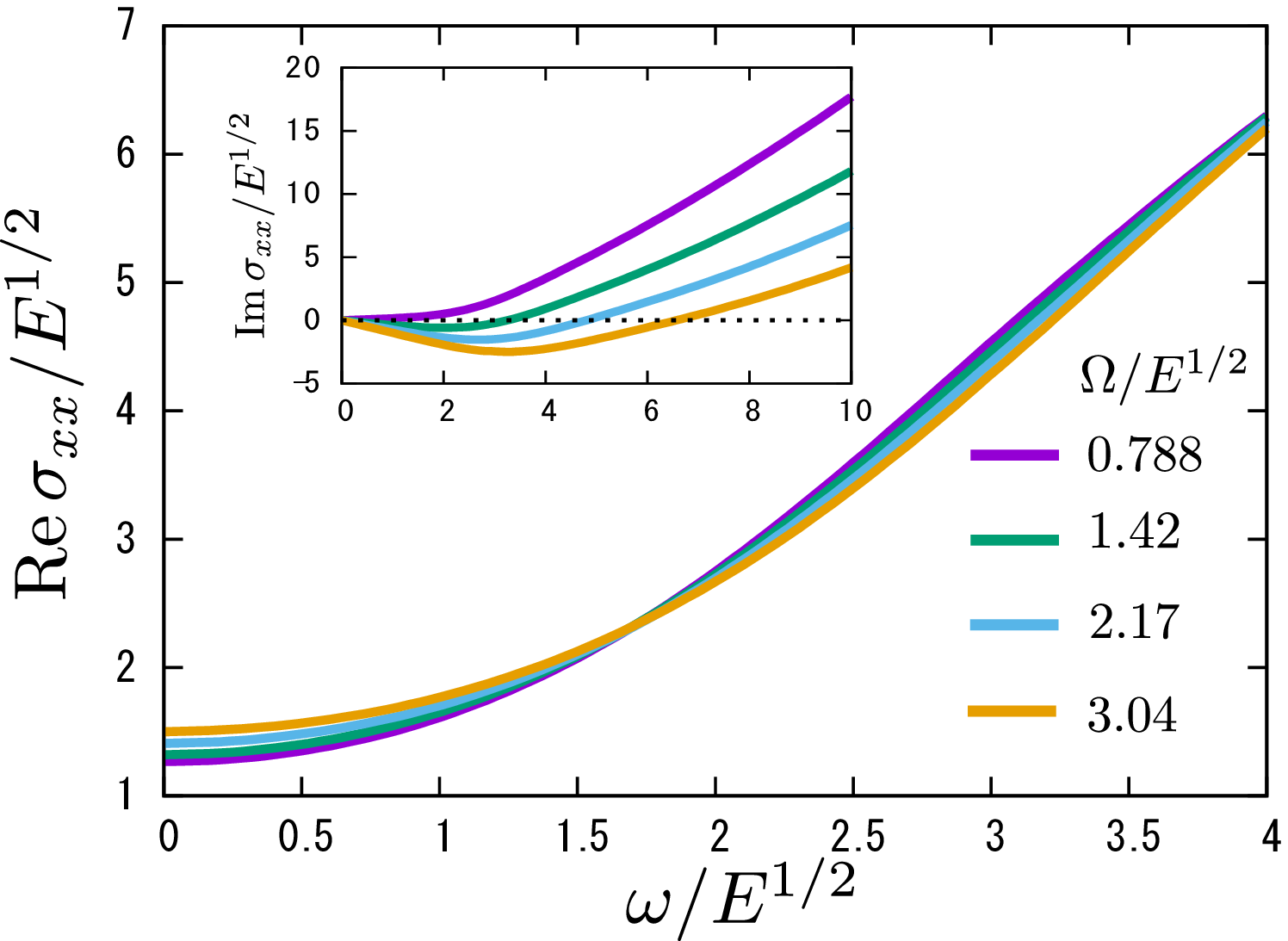}
  }
 \caption{
 The optical absorption spectrum against the probe AC frequency $\omega$.
 Curves in the left (right) figure correspond to several $E$ ($\Omega$)
 for a fixed $\Omega$ ($E$).
}
\label{Sxx_AC}
\end{figure}

In Fig.~\ref{Sxy_AC}, we show the optical Hall conductivities $\sigma_{xy}$
against the probe AC frequency $\omega$.
Curves in the left (right) figure correspond to several $E$ ($\Omega$)
for a fixed $\Omega$ ($E$).
In the DC limit $\omega\to 0$, we can check that
$\textrm{Re}\,\sigma_{xy}$ approaches the DC Hall coefficient obtained in Fig.~\ref{syx}.
In the both figures,  $\textrm{Re}\,\sigma_{xy}$ have peaks in the negative region. 
The peaks are amplified and their positions shift to higher frequency as
$E$ ($\Omega$) increases in the left (right) figure.
This is consistent with the condensed matter calculation in Ref.~\cite{OkaAoki,Dehghani:20152}. 
The calculation of the optical Hall response done in Ref.~\cite{Dehghani:20152} 
for a Floquet topological Hall state coupled to a phonon bath shows a striking resemblance with the 
holographic result Fig.~\ref{Sxy_AC}(a). 
They ($\textrm{Re}\,\sigma_{xy}$) both start from a flat region for small $\omega$
and goes negative at an intermediate frequency, finally approaching zero at $\omega\to \infty$. 
In the current system, the gluons not only mediates interaction 
but also act as a heat bath. This is  because we take the large $N_c$ limit
and the gluons are always in the equilibrium zero-temperature state. 
In this sense, the NESS obtained here seems to have similar properties 
as that of  Ref.~\cite{Dehghani:20152}. 
In Fig.~\ref{Sxx_AC}, we also show the optical absorption spectrum $\sigma_{xx}$
against $\omega$. $\textrm{Re}\,\sigma_{xy}$ is an increasing function of $E$ and $\Omega$.

The other conductivities $\bm{\sigma}^\pm$ are studied in appendix~\ref{app:other}.

\section{Conclusion and discussion}

Weyl semimetal can be created from Dirac semimetals by applying rotating electric fields. 
In this paper,
we have considered a similar set up in a strong coupling limit using AdS/CFT correspondence.
As a toy model of the strongly coupled field theory,
we have focused on $\mathcal{N}=2$ $SU(N_c)$ supersymmetric QCD at large
$N_c$ and at strong 't Hooft coupling, which is realized by the probe
D7-brane in the AdS$_5\times S^5$ spacetime.
We have computed the electric current, Joule heating and temperature
induced by the external rotating electric field~(\ref{rotEbc}).
They can be computed not only for the linear response regime
but for nonlinear regime thanks to the AdS/CFT correspondence.
We have found that they are an increasing function of $\Omega$ ($E$) for
a fixed $E$ ($\Omega$)
and nicely interpolates Ohm's law at high frequency to conformal DC conductivity at low frequency.
We have also studied its linear response. 
When the background rotating electric field is applied to the system,
the Hall currents appear as linear response to weak DC and AC probe electric fields.
The DC and AC Hall effects are expected for Weyl semimetal with free electron picture.
Here we have shown an example of strongly correlated fermions exhibiting the Hall effects.
We also find frequency mixed response currents, i.e., a heterodyning effect, 
characteristic to periodically driven Floquet systems.

We hope that physical quantities computed in this paper,
such as the effective temperature, Joule heating, electric current and conductivities,
capture universal behaviors of strongly coupled NESS created by the rotating electric field.
There is a steady energy flow within the degrees of freedom: Initially the 
quarks are coherently excited by the electric field. Then, Joule heating will take place and the 
effective temperature will rise. However, the system reaches a steady state 
since the quarks are coupled with the gluons which is kept in a
equilibrium zero-temperature state because we have taken the large $N_c$ limit. 
If we could work with a finite $N_c$, 
it is likely that the 
gluons will also heat up and eventually the system may drift to an infinite temperature state \cite{Lazarides:2014}.

In a field theory side,
computations of the transport properties of a NESS are usually difficult because 
we need to know its non-equilibrium distribution of electrons.
Once we assume that it is approximated by the equilibrium distribution function,
the DC Hall conductivity of a Floquet Weyl semimetal is known to be 
proportional to the separation of Weyl points in the momentum space~\cite{Wang:2014,Burkov}.
For the weak coupling limit, as studied in section~\ref{sec:weakcoupling},
the separation of Weyl points is given by $\Delta p \sim E^2/\Omega^3$ for $E/\Omega^2\ll 1$.
Thus, the DC Hall conductivity in weak coupling would be given by
$\sigma_{xy}/\Omega\sim (E/\Omega^2)^\alpha$ with $\alpha=2$ for $E/\Omega^2\ll 1$.
For strong coupling limit, on the other hand, fitting the plot in Fig.~\ref{syxE}, we obtain $\alpha\simeq 0.7$.
This discrepancy suggests that the power $\alpha$ would change at the strongly coupled NESS.

Our analysis opens up a way to analyze nonlinear effects of oscillatory electric field with a circular 
polarization. In particular, we are interested in how a deconfinement transition can be triggered 
by tuning the frequency $\Omega$ of the external electric field. It would provide
an interesting dynamical phase diagram of QCD-like gauge theories --- confinement/deconfinement
phase diagram with the axes of the amplitude/frequency of the external electric field.
Such kind of study would need holographic approach with an
intense and oscillatory electric field, and our method presented here can definitely help. We would like
to report on it soon \cite{soon}.

\subsection*{Acknowledgments}
We would like to thank Leda Bucciantini, Sthitadhi Roy, Sota Kitamura, Kenichi Asano,  Hideo Aoki, and Ryo Shimano for valuable discussions.
The work of K.~H. was supported in part by JSPS KAKENHI Grant Number 15H03658 and 15K13483.
The work of S.~K. was supported in part by JSPS KAKENHI Grant Number JP16K17704.
The work of K.~M. was supported by JSPS KAKENHI Grant Number 15K17658.
The work of T.~O.  was supported by JSPS KAKENHI Grant Number 23740260 and the ImPact project (No. 2015-PM12-05-01) from JST.

\appendix

\section{Regular solution near the effective horizon}
\label{app:regsol}

We study the regular solution of Eq.~(\ref{beq})
near the effective horizon $\rho=\rho_c$.
For simplicity, we take the unit of $\Omega=1$ and set $\theta=0$ in this section.
We expand $b(\rho)$ near the horizon as
\begin{equation}
 b(\rho)=\rho_c^2 + p (\rho-\rho_c)+\cdots\ ,
  \label{breg_app}
\end{equation}
where $p$ is a complex constant which will be determined by the regularity.
Substituting the above expression into Eq.~(\ref{beq}), we obtain equation for $p$ from the leading term in $\rho-\rho_c$ as
\begin{equation}
\rho_c\, p^3-10 \rho_c\, p^2 p^\ast+\rho_c\, pp^\ast{}^2-4p^2-8\rho_c\, p^\ast-4=0\ ,\label{peq1}
\end{equation}
Multiplying (\ref{peq1}) by $p^*$ and taking its imaginary part, we obtain
\begin{equation}
 (p-p^\ast)(pp^\ast-2\rho_c\,p-2\rho_c\,p^\ast-1)=0\ .
\end{equation}
This yields two equations, 
\begin{equation}
 p^\ast=p\ ,\quad
 p^* = \frac{2\rho_c\,p+1}{p-2\rho_c}\ .
 \label{past_sol}
\end{equation}
The first equation implies that $p$ is a real value. 
In this case, the Ricci scalar with respect to the effective metric~(\ref{effmet}) becomes
\begin{equation}
 R(\gamma)\simeq \frac{\rho_c^2(p^4+6p^2-16\rho_c p +16\rho_c^2+1)}{8(1+p^2)(p-2\rho_c)^2(\rho-\rho_c)^2}\qquad
  (\rho\to \rho_c)\ .
\end{equation}
The naked singularity appears at $\rho=\rho_c$ when $p$ is a real value.
So, we consider the second equation in Eq.~(\ref{past_sol}).
Substituting it into Eq.~(\ref{peq1}) and dividing it by $p$,
we obtain 4th order equation for $p$ as
\begin{equation}
 \rho_c(9+32\rho_c^2)-4(1+2\rho_c^2)p+6\rho_c(1+8\rho_c^2)p^2-4(1+6\rho_c^2)p^3+\rho_c\, p^4=0\ .
\end{equation}
We have four solutions of this equation as 
\begin{equation}
 p=\left\{
\begin{aligned}
 \mathcal{R}_6&-\sqrt{\mathcal{R}_4\mathcal{R}_9}\pm
 i\sqrt{2\mathcal{R}_4(\sqrt{\mathcal{R}_4\mathcal{R}_9}-\mathcal{R}_6)} \\
 \mathcal{R}_6&+\sqrt{\mathcal{R}_4\mathcal{R}_9}\pm \sqrt{2\mathcal{R}_4(\sqrt{\mathcal{R}_4\mathcal{R}_9}+\mathcal{R}_6)}
\end{aligned}
\right. \ ,
\label{psols}
\end{equation}
where $\mathcal{R}_n$ is defined below Eq.~(\ref{horreg}).
The latter two solutions do not satisfy the original
equation~(\ref{peq1}) and they are fictional solutions.
Let us focus on the former two solutions, which are true solutions of
(\ref{peq1}).
When we take the positive signature of the former solutions, 
we find that the Joule heating is positive as in Fig.~\ref{JQE00}.
On the other hand, if we take the negative signature,
the solution is
replaced as $b\leftrightarrow b^\ast$, $E\leftrightarrow -E^\ast$ and $j\leftrightarrow j^\ast$.
Then, 
the Joule heating is replaced as $q\leftrightarrow -q$ and becomes negative.
Briefly speaking, this means time reversal solutions.
In the view of the effective geometry,
there is a energy flux from the white hole horizon.
In this paper,
we adopt the positive signature of the former solution in Eq.~(\ref{psols}). 

\section{Perturbation equations}
\label{app:pert}

The action for the perturbation of $b$ is obtained by replacing 
$b(t,\rho)$ with $b(\rho)+\delta b(t,\rho)$ in Eq.~(\ref{action_b}) and taking the second order in $\delta b$.
Here, $b(\rho)$ is the background solution satisfying Eq.~(\ref{beq}).
The equation of motion for $\delta b$ is given by Eq.~(\ref{vecbeq}).
Components of matrices $\bm{A}$, $\bm{B}$ and $\bm{C}$ are written as
\begin{equation}
\begin{split}
 &A_{11}=
 2 \Omega \rho\mathcal{L}_0^{-2}\{2  (b_1 b_1'+b_2 b_2')+\rho\} (\Omega^2  b_1 b_2+\rho^4 b_1' b_2')\ ,\\
 &A_{12}=
 -2 \Omega \rho \mathcal{L}_0^{-2} \{2  (b_1 b_1'+b_2 b_2')+\rho\}
 (-\Omega^2 b_2^2+ \rho^4 b_1'{}^2+\rho^4) \ ,\\
 &A_{21}= -(b_1 \leftrightarrow b_2 \textrm{ in } A_{12})\ ,\\
 &A_{22}=-A_{11}\ ,\\
 &B_{11}=
  \mathcal{L}_0^{-1}
 [2  \Omega^2 (b_1 b_1'- \rho) (b_1 b_1'+ b_2 b_2')
 +\Omega^2 (b_1^2+5 b_2^2)-6 \rho^4 b_1'{}^2-\rho^4]\ ,\\
 &B_{12}=
 2 \mathcal{L}_0^{-1}[\Omega^2 b_1' b_2 (b_1 b_1'+b_2 b_2')
 +\Omega^2 \{\rho(b_1 b_2'- b_1' b_2)-2  b_1 b_2\}-3 \rho^4 b_1' b_2']\ ,\\
 &B_{21}=(b_1 \leftrightarrow b_2 \textrm{ in } B_{12})\ ,\\
 &B_{22}=(b_1 \leftrightarrow b_2 \textrm{ in } B_{11})\ ,\\
&C_{11}=-\Omega^2 \rho \mathcal{L}_0^{-2}
 [
 -2 \Omega^2 b_1' b_2 (b_1 b_2'-b_1' b_2) (b_1 b_1'+b_2 b_2')\\
 &\hspace{1cm}
 + \Omega^2 \rho (b_1 b_1'+b_1 b_2'-b_1' b_2+b_2 b_2') (b_1 b_1'-b_1 b_2'+b_1' b_2+b_2 b_2')\\
 &\hspace{2cm}
 +4  \Omega^2 b_2 (b_1^2 b_2'-2 b_1 b_1' b_2-b_2^2 b_2')
 +2 \rho^4 b_1' (2 b_1 b_1'{}^2+3 b_1 b_2'{}^2-b_1' b_2 b_2')\\
 &\hspace{3cm}
 + \Omega^2 \rho (b_1^2-b_2^2)
+ \rho^5 (b_1'{}^2-b_2'{}^2) 
 +4  \rho^4 (b_1 b_1'+b_2 b_2')+\rho^5
 ]\ ,\\
 &C_{12}=
 2  \Omega^2 \rho \mathcal{L}_0^{-2} [
 - \Omega^2 b_1 b_1' (b_1 b_2'-b_1' b_2) (b_1 b_1'+b_2 b_2')\\
 &\hspace{2cm}
 + \Omega^2 \rho (b_1 b_2'-b_1' b_2) (b_1 b_1'+b_2 b_2')
 +2  \Omega^2 b_1 (b_1^2 b_2'-2 b_1 b_1' b_2-b_2^2 b_2')\\
 &\hspace{4cm}
 + b_1' \rho^4 (b_1 b_1' b_2'-3 b_1'{}^2 b_2-2 b_2 b_2'{}^2)
 - \Omega^2 \rho b_1 b_2 \\
 &\hspace{6cm}
 -2 \rho^4 (b_1 b_2'-b_1' b_2)
 -\rho^5 b_1' b_2' 
 ]\ ,\\
 &C_{21}=(b_1 \leftrightarrow b_2 \textrm{ in } C_{12})\ ,\\
 &C_{22}=(b_1 \leftrightarrow b_2 \textrm{ in } C_{11})\ ,
\end{split}
\end{equation}
where $b_1=\textrm{Re}\, b$ and $b_2=\textrm{Im}\, b$.
We have eliminated $d^2b/d\rho^2$ using Eq.~(\ref{beq}) in the above expressions.

Now, we study the asymptotic behavior of $\delta \vec{b}$ 
at the horizon $\rho=\rho_c$.
The asymptotic form of the background solution $b(\rho)$ is given by Eq.~(\ref{breg_app}).
Substituting Eq.~(\ref{breg_app}) into $\bm{A}, \bm{B}$ and
$\bm{C}$, we obtain
\begin{equation}
\begin{split}
&\bm{A}_0\equiv \bm{A}(\rho=\rho_c)=
\begin{pmatrix}
 2p_1(2 \rho_c p_1+1)/p_2 &
 -2(1+p_1^2)(2\rho_c p_1+1)/p_2^2\\
2(2\rho_c p_1+1) &
-2p_1(2 \rho_c p_1+1)/p_2
\end{pmatrix}\ ,\\
&\bm{B}_0\equiv \bm{B}(\rho=\rho_c)=
\begin{pmatrix}
 -2p_1(2 \rho_c p_1+1)/p_2 &
 -2(3\rho_c p_1-1)\\
-2(2\rho_c p_1+1) &
-2(3\rho_c p_2^2-2\rho_c +p_1)/p_2
\end{pmatrix}\ ,\\
&\bm{C}_0\equiv \bm{C}(\rho=\rho_c)=
\begin{pmatrix}
-2(1+2\rho_c p_1+p_1^2+2\rho_c p_1^3-p_2^2+3\rho_c p_1 p_2^2)/p_2^2 &
0 \\
-2(-2\rho_c+2p_1+2\rho_c p_1^2+3\rho_c p_2^2)/p_2 &
0
\end{pmatrix}\ ,
\end{split}
\end{equation}
where $p_1=\textrm{Re}\, p$ and $p_2=\textrm{Im}\, p$.
We have not used the explicit expression for $p$ in the above expressions.
Substituting the explicit expression of $p$ (the positive signature of the former solution in Eq.~(\ref{psols}))
into the above expressions, 
we have $\bm{B}_0=-\bm{A}_0$ and
$\bm{C}_0=0$.
Eventually, near the horizon $\rho=\rho_c$, 
the perturbation equation becomes
\begin{equation}
 (\partial_\tau^2-\partial_{\rho_\ast}^2)\delta \vec{b}
+\bm{A}_0(\partial_\tau -\partial_{\rho_\ast})\delta \vec{b}=0\ .
\end{equation}
Therefore, the asymptotic solution is
\begin{equation}
 \delta \vec{b}\simeq \vec{f}(t+\rho_\ast)
+e^{\bm{A}_0 \rho_\ast}\, \vec{g}(t-\rho_\ast)
\qquad (\rho\to \rho_c)\ .
\end{equation}
We will impose $\vec{g}=0$ from the ingoing wave boundary condition.

\section{Relations in conductivity matrices}
\label{sym_cund}

We give a proof of the relations in conductivities~(\ref{sigmasym}).
Since the perturbation equations~(\ref{beta_eq}) are linear,
$\beta_\pm^{(0)}$ and $\beta_\pm^{(2)}$ defined in Eq.~(\ref{beta_inf}) are linearly related as
\begin{equation}
 \vec{\beta}_\pm^{(2)}=\bm{X}_\pm \vec{\beta}_\pm^{(0)}=-\frac{i}{\omega} \bm{X}_\pm \bm{M}_\pm \vec{\varepsilon}_\omega\ ,
\end{equation}
where $\bm{X}_\pm$ is a $2\times 2$ complex matrix. At the last equality, we used Eq.~(\ref{betainf}).
Substituting the above expressions into Eq.~(\ref{j_beta2}), we obtain
\begin{multline}
 \delta\vec{j}\propto \big [
  e^{-i\omega t}(\bm{M}_+\bm{X}_+ \bm{M}_+ + \bm{M}_-\bm{X}_- \bm{M}_-)\\
  +e^{-i(\omega+2\Omega) t}\bm{M}_-\bm{X}_+ \bm{M}_+
  +e^{-i(\omega-2\Omega) t}\bm{M}_+\bm{X}_- \bm{M}_- \big]\vec{\varepsilon}_\omega\ .
\label{j_beta3}
\end{multline}
Therefore, conductivities are written as
\begin{equation}
 \bm{\sigma}\propto \bm{M}_+\bm{X}_+ \bm{M}_+ + \bm{M}_-\bm{X}_- \bm{M}_-\ ,\qquad
 \bm{\sigma}^\pm\propto \bm{M}_\mp \bm{X}_\pm \bm{M}_\pm\ .
\end{equation}
We will denote the components of $\bm{X}_\pm$ as $x_{ij}^\pm$ ($i=1,2$). By the explicit calculation of
matrix multiplications, we obtain
\begin{equation}
\begin{split}
&\bm{\sigma}\propto
  (x^+_{11}+ix^+_{21}-ix^+_{12}+x^+_{22})\bm{M}_+
 +(x^-_{11}-ix^-_{21}+ix^-_{12}+x^-_{22})\bm{M}_-\ ,\\
 &\bm{\sigma}^\pm \propto
 (x^\pm_{11}\mp ix^\pm_{21}\mp ix^\pm_{12}-x^\pm_{22})
 \begin{pmatrix}
  1 & \pm i \\
  \pm i & -1
 \end{pmatrix}\ .
\end{split}
\end{equation}
From the above expressions, we can find the relations in Eq.~(\ref{sigmasym}).

\section{Numerical method for the time domain approach}
\label{app:num}

\begin{figure}
\begin{center}
\includegraphics[scale=0.4]{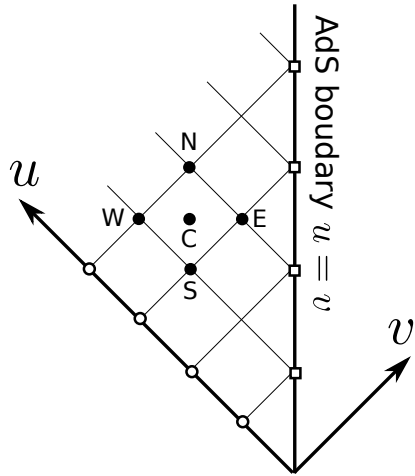}
\end{center}
\caption{
World volume of D7-brane. The mesh for numerical calculation is taken along double null coordinates $u$ and $v$.
}
 \label{timedomain_num}
\end{figure}

In this section, we explain how to solve the perturbation equation~(\ref{vecbeq}) in the time domain.
We introduce double null coordinates as $u=(\tau+\rho_\ast)/2$ and $v=(\tau-\rho_\ast)/2$.
Then, the perturbation equation becomes
\begin{equation}
 \left[\partial_u \partial_v
+\frac{1}{2}(\bm{A}+\bm{B})\partial_u
+\frac{1}{2}(\bm{A}-\bm{B})\partial_v
+\bm{C}\right]\delta \vec{b}=0\ .
\label{vecbequv}
\end{equation}
We consider a numerical mesh along $(u,v)$-coordinates as in Fig.~\ref{timedomain_num}.
At points shown by white circles ($\circ$), we give a trivial initial condition: $\delta \vec{b}=0$.
At points shown by white squares ({\tiny{$\Box$}}), we impose the boundary condition in Eq.~(\ref{dbbdry}).
To determine the solution inside the numerical domain, we discretize the derivatives at point C in the figure as
\begin{equation}
\begin{split}
 &\partial_u \partial_v \delta \vec{b}|_\mathrm{C} = (\delta \vec{b}_\mathrm{N}-\delta \vec{b}_\mathrm{E}-\delta \vec{b}_\mathrm{W}+\delta \vec{b}_\mathrm{S})/h^2\ ,\\
 &\partial_u \delta \vec{b}|_\mathrm{C} = (\delta \vec{b}_\mathrm{N}-\delta \vec{b}_\mathrm{E}+\delta \vec{b}_\mathrm{W}-\delta \vec{b}_\mathrm{S})/(2h)\ ,\\
 &\partial_v \delta \vec{b}|_\mathrm{C} = (\delta \vec{b}_\mathrm{N}+\delta \vec{b}_\mathrm{E}-\delta \vec{b}_\mathrm{W}-\delta \vec{b}_\mathrm{S})/(2h)\ ,\\
 &\delta \vec{b}|_\mathrm{C} = (\delta \vec{b}_\mathrm{E}+\delta \vec{b}_\mathrm{W})/2\ ,
\end{split}
\end{equation}
where $h$ is the mesh size and $\delta \vec{b}_i$ ($i=\mathrm{N}, \mathrm{E}, \mathrm{W}, \mathrm{S}$) represents numerical value of $\delta \vec{b}$ at point $i$.
They have second order accuracy.
Substituting the above expressions into Eq.~(\ref{vecbequv}),
we can express $\delta \vec{b}_\mathrm{N}$ by $\delta \vec{b}_\mathrm{E}$, $\delta \vec{b}_\mathrm{W}$ and  $\delta \vec{b}_\mathrm{S}$.
We evaluate $\rho$-dependent matrices $\bm{A},\bm{B}$ and $\bm{C}$ at the point C.
Using the discretized evolution equation sequentially, we can determine the solution in the whole numerical domain.

\section{Other conductivities}
\label{app:other}

In section~\ref{sec:sigmas}, we have showed that
three modes are induced in the electric current
when we apply monochromatic probe AC electric field to the holographic Floquet state.
We have defined three kinds of conductivities $\bm{\sigma}$ and
$\bm{\sigma}^\pm$ as in Eq.~(\ref{sigma_def}).
In sections~\ref{sec:DCHall} and \ref{sec:ACHall},
we have focused only on $\bm{\sigma}$.
Here, we study the other parts of conductivities, $\sigma^\pm$.

\begin{figure}
  \centering
  \subfigure[Varying $E$ for fixed $\Omega$]
  {\includegraphics[scale=0.45]{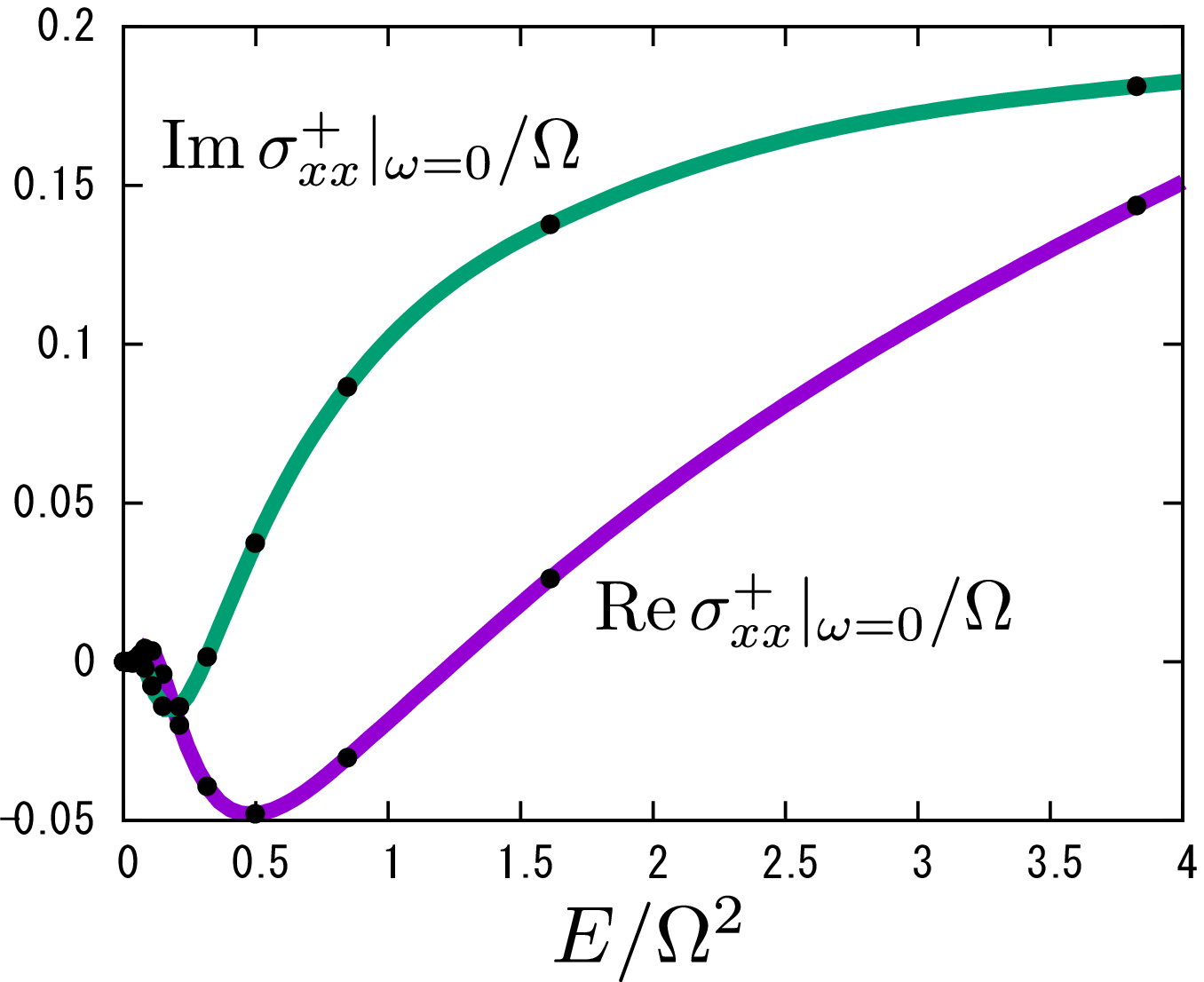}\label{sxxE}
   }
  \subfigure[Varying $\Omega$ for fixed $E$]
  {\includegraphics[scale=0.45]{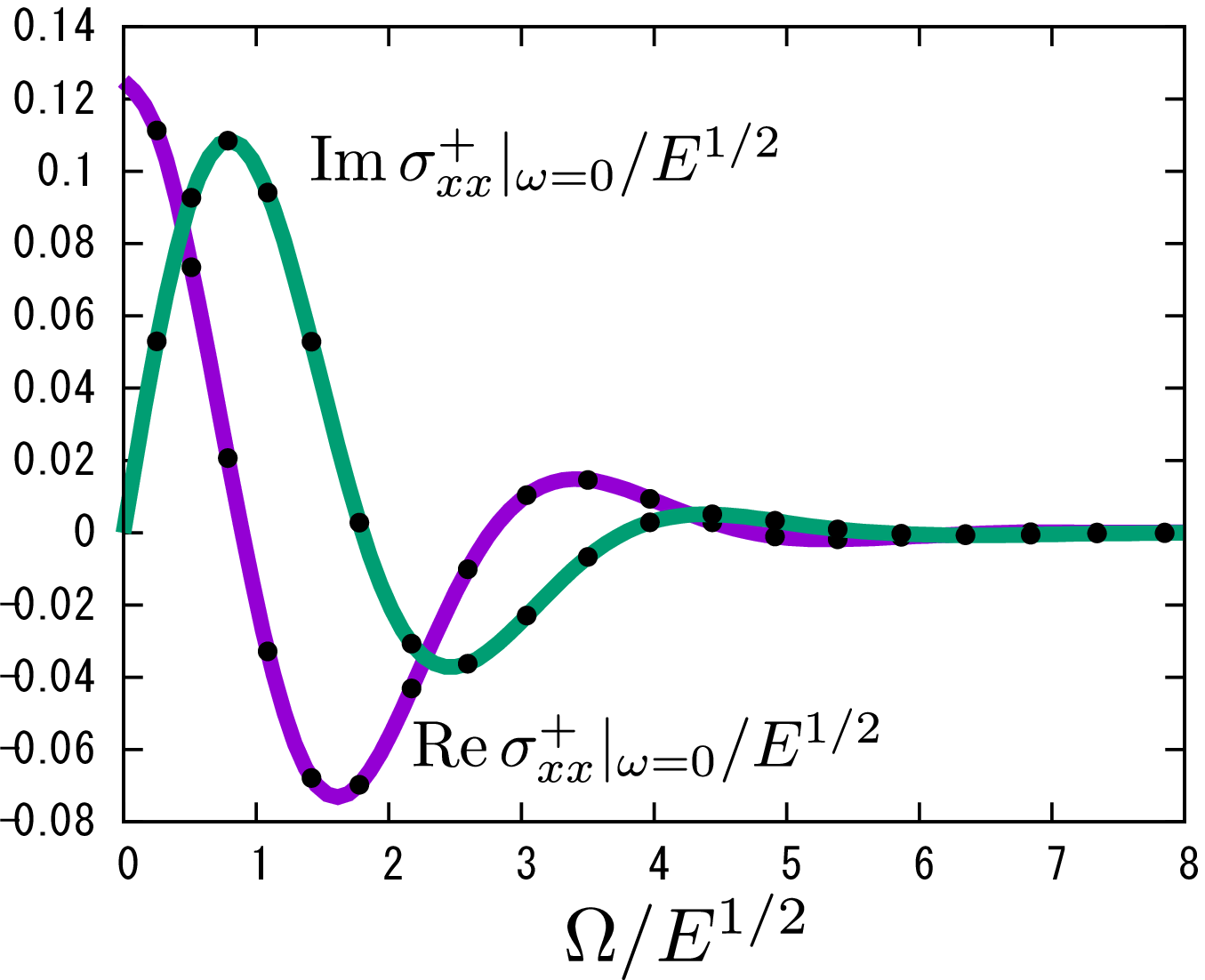}\label{sxxOm}
  }
  \caption{
Conductivity $\sigma_{xx}^+$ against probe DC electric field.
}
\label{spxx}
\end{figure}

Firstly, we consider the probe DC electric field $\omega=0$.
In this case, $\bm{\sigma}^-|_{\omega=0}$ is given by the  complex conjugate of
$\bm{\sigma}^+|_{\omega=0}$. 
In Fig.~\ref{spxx}, we show $\sigma_{xx}^+|_{\omega=0}$ against background parameters:
(a) $\sigma_{xx}^+|_{\omega=0}/\Omega$ vs $E/\Omega^2$, 
and (b) $\sigma_{xx}^+|_{\omega=0}/E^{1/2}$ vs $\Omega/E^{1/2}$.
When we fix the background electric field $E$,
the conductivity $\bm{\sigma}^+|_{\omega=0}$ oscillates as a function of $\Omega$ for $\Omega\lesssim 4 E^{1/2}$ but 
it is suppressed for $\Omega\gtrsim 4 E^{1/2}$.
The $\sigma_+$ expresses the oscillating component of the electric current 
as one can see in Fig.~\ref{dJ}.
It follows that, for $\Omega\gtrsim 4 E^{1/2}$,
almost stationary current is induced by the probe DC electric field.

\begin{figure}
  \centering
  \subfigure[Varying $E$ for fixed $\Omega$]
  {\includegraphics[scale=0.4]{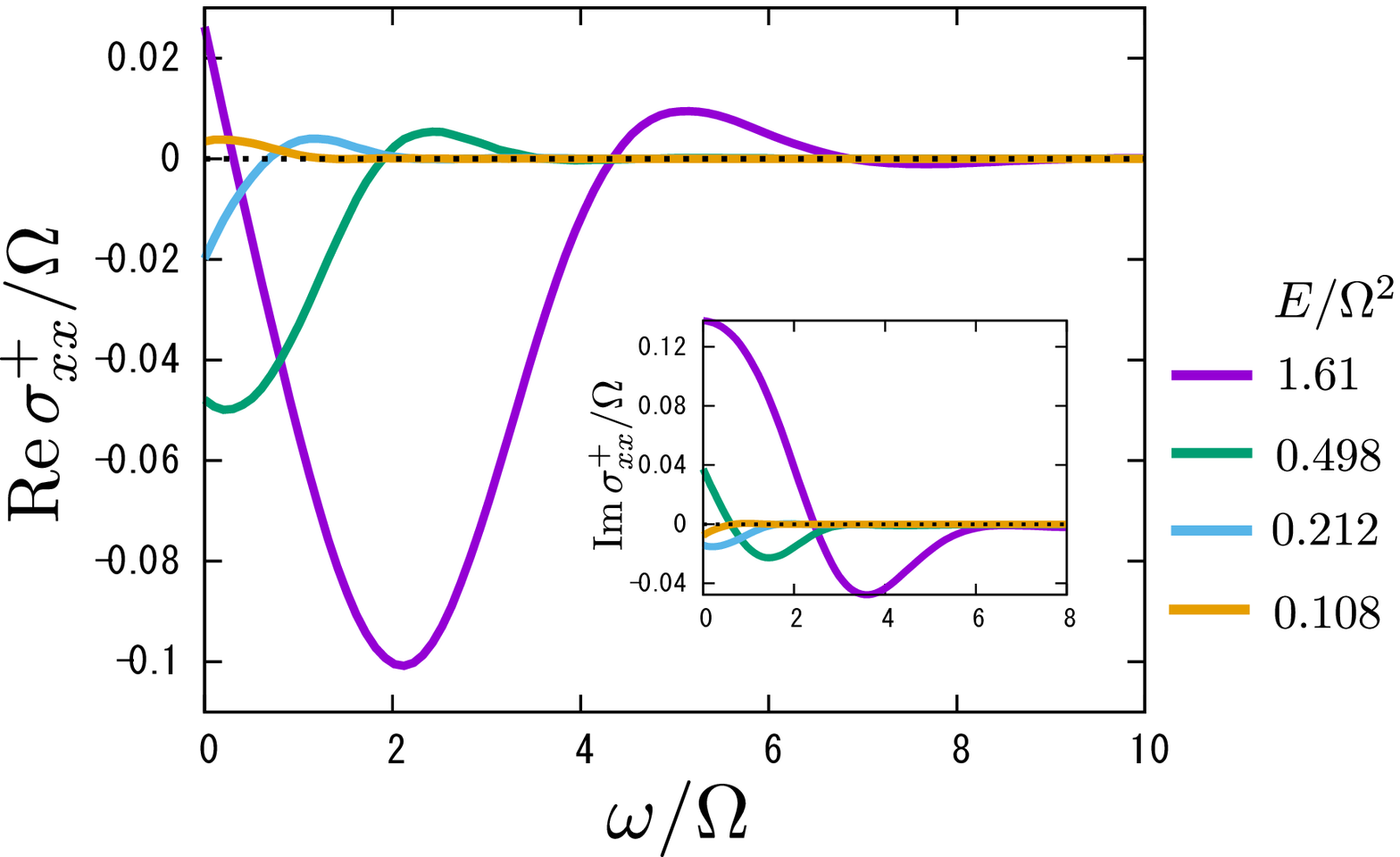}
   }
  \subfigure[Varying $\Omega$ for fixed $E$]
  {\includegraphics[scale=0.4]{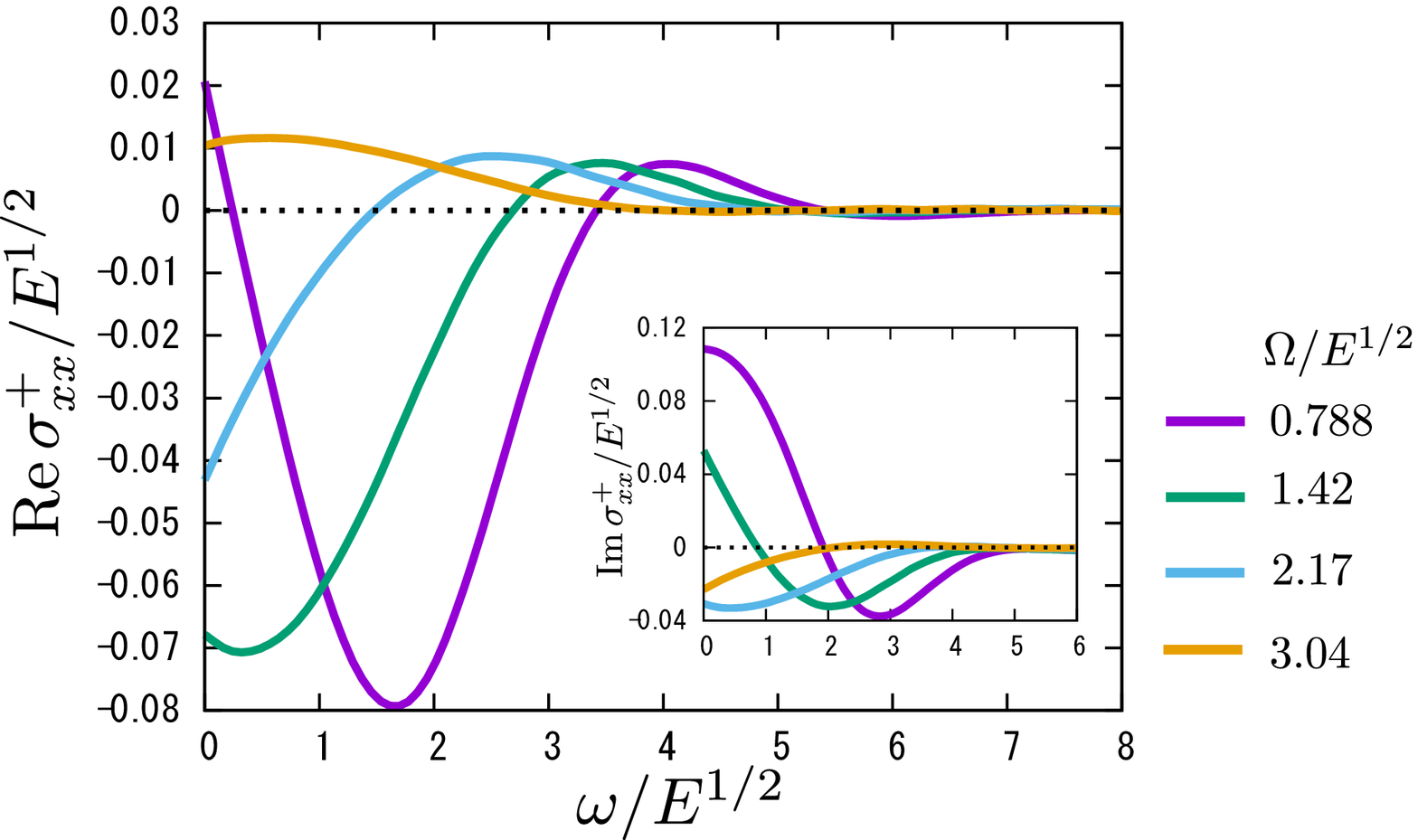}
  }
  \caption{
Conductivity $\sigma^+_{xx}$ against probe AC electric field.
}
\label{spxx_ac}
\end{figure}

\begin{figure}
  \centering
  \subfigure[Varying $E$ for fixed $\Omega$]
  {\includegraphics[scale=0.4]{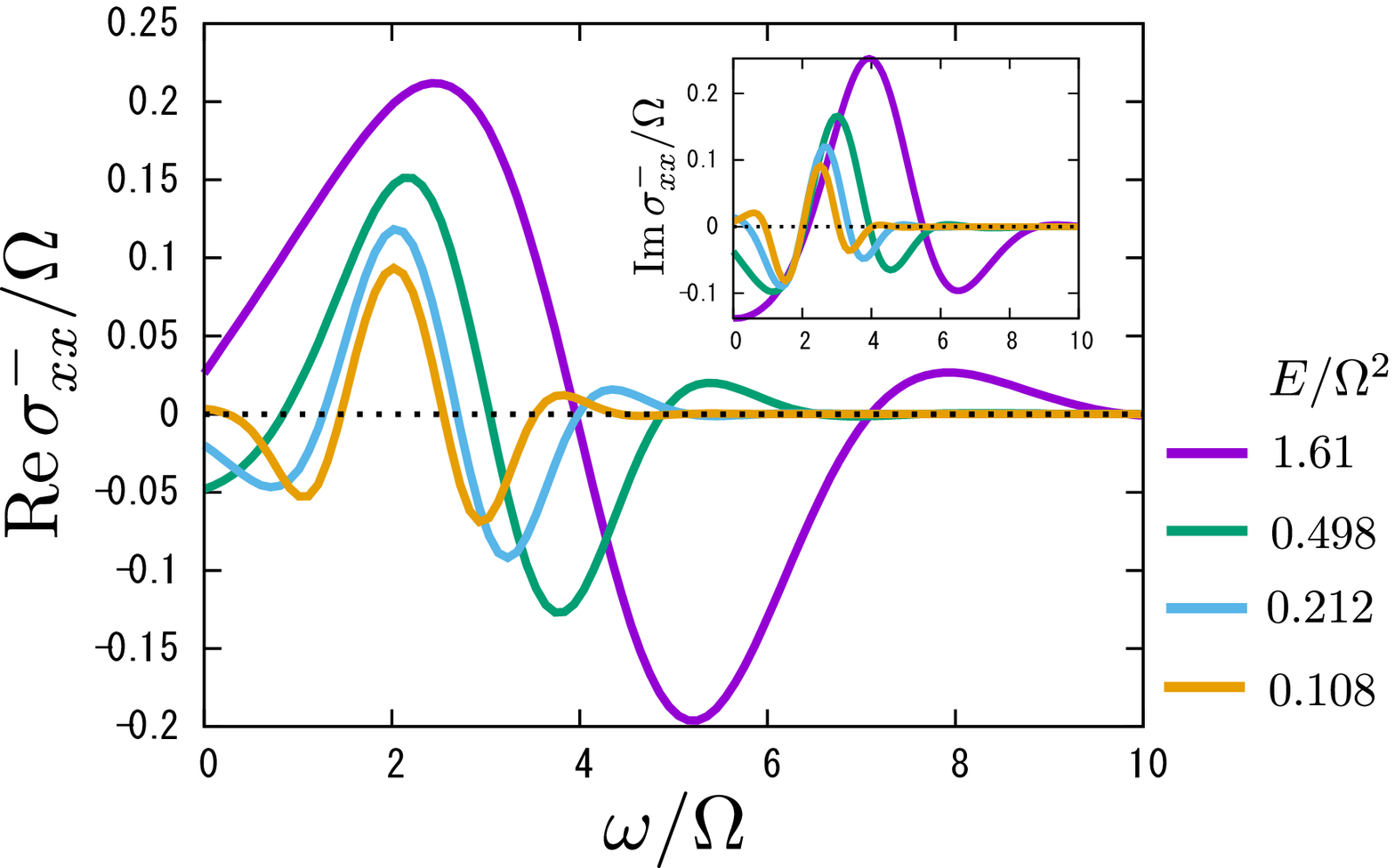}
   }
  \subfigure[Varying $\Omega$ for fixed $E$]
  {\includegraphics[scale=0.4]{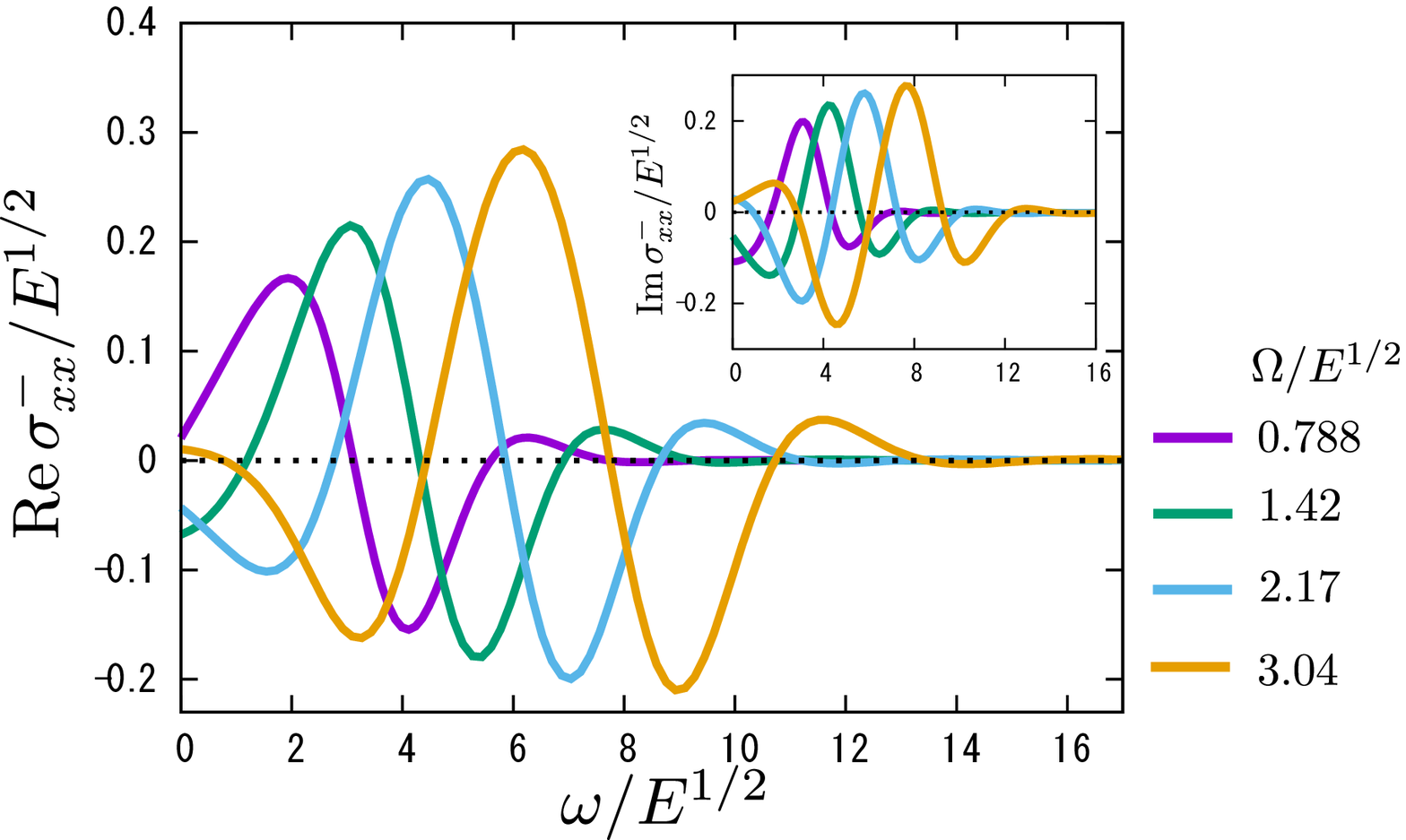}
  }
 \caption{
 Conductivity $\sigma^-_{xx}$ against probe AC electric field.
}
\label{smxx_ac}
\end{figure}

For the probe AC electric field, the conductivities $\sigma_{xx}^+$ and $\sigma_{xx}^-$ are shown in 
Figs.~\ref{spxx_ac} and \ref{smxx_ac}, respectively.
They are all suppressed for large $\omega$.
For a fixed $\Omega$, $\sigma_{xx}^\pm$ are amplified as $E$ increases.
When we fix $E$, $\sigma_{xx}^+$ is getting small as $\Omega$ increases.
On the other hand,
$\sigma_{xx}^-$ becomes large and 
the position of the peak is shifted to higher frequency as $\Omega$ increases.

\begin{thebibliography}{99}


\bibitem{Maldacena:1997re} 
  J.~M.~Maldacena,
  ``The Large N limit of superconformal field theories and supergravity,''
  Adv.\ Theor.\ Math.\ Phys.\  {\bf 2}, 231 (1998)
  [hep-th/9711200].

\bibitem{Gubser:1998bc} 
  S.~S.~Gubser, I.~R.~Klebanov and A.~M.~Polyakov,
  ``Gauge theory correlators from noncritical string theory,''
	Phys.\ Lett.\ B {\bf 428}, 105 (1998)
  [hep-th/9802109].
 
\bibitem{Witten:1998qj} 
  E.~Witten,
  ``Anti-de Sitter space and holography,''
  Adv.\ Theor.\ Math.\ Phys.\  {\bf 2}, 253 (1998)
  [hep-th/9802150].



\bibitem{Hartnoll:2008}
S. A. Hartnoll, C. P. Herzog, G. T.  Horowitz,
  Phys.\ Rev.\ Lett. {\bf 101}, 31601 (2008).
  
\bibitem{Cubrovic:2009}
M. Cubrovi\'c, Jan Zaanen, K. Schalm,
Science {\bf 325}, 439 (2009).

\bibitem{Huijse:2012}
L. Huijse, S. Sachdev, B. Swingle,
  Phys.\ Rev.\ B {\bf 85}, 035121 (2012).



\bibitem{Ebihara:2016} 
  S.~Ebihara, K.~Fukushima, T.~Oka,
  ``Chiral pumping effect induced by rotating electric fields,''
 Phys.\ Rev.\ B\  {\bf 93}, 155107 (2016)
  [hep-th/1203.4908].

\bibitem{Sambe:1973}
H.~Sambe,
  Phys.\ Rev.\ A {\bf 7}, 2203 (1973).


\bibitem{Lindner11} 
N. H. Lindner, G. Refael, and V. Galitski, 
  Nat.\ Phys. {\bf 7}, 490 (2011).	

\bibitem{Oka09} 
T. Oka and H. Aoki, 
  Phys.\ Rev.\ B {\bf 79}, 081406 (2009).
  
\bibitem{Kitagawa10} 
T. Kitagawa, M. S. Rudner, E. Berg, and E. Demler, 
  Phys.\ Rev.\ A {\bf 82}, 0033429 (2010).

\bibitem{Kitagawa11} 
T. Kitagawa, T. Oka, A. Brataas, L. Fu, and E. Demler, 
  Phys.\ Rev.\ B {\bf 84}, 235108 (2011).

\bibitem{Haldane1988} 
F. D. M. Haldane,
  ``Model for a Quantum Hall Effect without Landau Levels: Condensed-Matter Realization of the "Parity Anomaly",''
  Phys.\ Rev.\ Lett.\  {\bf 61}, 2015 (1988).

\bibitem{Jotzu14}
G. Jotzu, M. Messer, R\'emi Desbuquois, M. Lebrat, T. Uehlinger, 
D. Greif, and T. Esslinger, Nature {\bf 515}, 237 (2014).

\bibitem{Wang13}  
Y. H. Wang, H. Steinberg, P. Jarillo-Herrero, and N. Gedik, 
Science, {\bf 342}, 453 (2013). 

\bibitem{Nielsen:1983} 
H.~B.~Nielsen, M.~Ninomiya,
  ``The Adler-Bell-Jackiw anomaly and Weyl fermions in a crystal,''
 Phys.\ Lett. \ {\bf 130B},  389 (1983).

\bibitem{Wang:2014} 
  R.~Wang, B.~Wang, R.~Shen, L.~Sheng and D. Y.~Xing
  ``Floquet Weyl semimetal induced by topological phase transitions,''
 EPL {\bf 105}, 17004 (2014).

\bibitem{XiaoXiao:2016}
X.-X. Zhang, T. T. Ong, N. Nagaosa,
"Theory of photoinduced Floquet Weyl semimetal phases,"
arXiv:1607.05941v1.

\bibitem{Hubener:2016} 
H. H\"ubener, M. A. Sentef, U. de Giovannini, A. F. Kemper, A. Rubio, 
  ``Creating stable Floquet-Weyl semimetals by laser-driving of 3D Dirac materials,'' 
  arXiv:1604.03399v3.

\bibitem{Oka-new}
L. Bucciantini, S. Roy, S. Kitamura, and T. Oka
T.~Oka and L.~Bucciantini, to appear.


\bibitem{Kohn:2014} 
W. Kohn,
  ``Periodic thermodynamics,''
J. Stat. Phys.  {\bf 103}, 417 (2014).

\bibitem{Lazarides:2014} 
A. Lazarides, A. Das, R. Moessner,
  ``Periodic thermodynamics of isolated systems,''
  Phys.\ Rev.\ Lett.\  {\bf 111}, 150401 (2014).
    
\bibitem{Lazarides:20142} 
A. Lazarides, A. Das, R. Moessner,
  Phys.\ Rev.\ E\  {\bf 90}, 012110 (2014).
  
\bibitem{DAlessio:2014} 
  L. D’Alessio and M. Rigol,
  Phys.\ Rev.\ X\  {\bf 4}, 041048  (2014).

\bibitem{Dehghani:2014}
H.~Dehghani, T.~Oka and A.~Mitra Phys. Rev. {\bf B 90} 195429 (2014).


\bibitem{Dehghani:20152}
H.~Dehghani  and A.~Mitra 
``Optical Hall conductivity of a Floquet topological insulator,''
Phys. Rev. {\bf B 92} 165111 (2015).

\bibitem{Karch}
A.~Karch and E.~Katz, 
``Adding flavor to AdS / CFT,'' 
  JHEP {\bf 0206} 043 (2002), [arXiv:hep-th/0205236].
  
\bibitem{Karch:2007pd} 
  A.~Karch and A.~O'Bannon,
  ``Metallic AdS/CFT,''
  JHEP {\bf 0709}, 024 (2007)
  [arXiv:0705.3870 [hep-th]].

\bibitem{Albash:2007bq} 
  T.~Albash, V.~G.~Filev, C.~V.~Johnson and A.~Kundu,
  ``Quarks in an external electric field in finite temperature large N gauge theory,''
  JHEP {\bf 0808}, 092 (2008)
  [arXiv:0709.1554 [hep-th]].

\bibitem{Erdmenger:2007bn} 
  J.~Erdmenger, R.~Meyer and J.~P.~Shock,
  ``AdS/CFT with flavour in electric and magnetic Kalb-Ramond fields,''
  JHEP {\bf 0712}, 091 (2007)
  [arXiv:0709.1551 [hep-th]].

\bibitem{Hashimoto:2013mua} 
  K.~Hashimoto and T.~Oka,
  ``Vacuum Instability in Electric Fields via AdS/CFT: Euler-Heisenberg Lagrangian and Planckian Thermalization,''
  JHEP {\bf 1310}, 116 (2013)
  [arXiv:1307.7423].
  
\bibitem{Hashimoto:2014dza} 
  K.~Hashimoto, T.~Oka and A.~Sonoda,
  ``Magnetic instability in AdS/CFT : Schwinger effect and Euler-Heisenberg Lagrangian of Supersymmetric QCD,''
To be published in JHEP
  [arXiv:1403.6336].

\bibitem{Hashimoto:2014yya} 
  K.~Hashimoto, T.~Oka and A.~Sonoda,
  ``Electromagnetic instability in holographic QCD,''
  JHEP {\bf 1506}, 001 (2015)
  [arXiv:1412.4254 [hep-th]].

\cite{Hashimoto:2014yza}
\bibitem{Hashimoto:2014yza} 
  K.~Hashimoto, S.~Kinoshita, K.~Murata and T.~Oka,
  ``Electric Field Quench in AdS/CFT,''
  JHEP {\bf 1409}, 126 (2014)
  [arXiv:1407.0798 [hep-th]].

\bibitem{Hashimoto:2014xta} 
  K.~Hashimoto, S.~Kinoshita, K.~Murata and T.~Oka,
  ``Turbulent meson condensation in quark deconfinement,''
  Phys.\ Lett.\ B {\bf 746}, 311 (2015)
  [arXiv:1408.6293 [hep-th]].

\bibitem{Hashimoto:2014dda} 
  K.~Hashimoto, S.~Kinoshita, K.~Murata and T.~Oka,
  ``Meson turbulence at quark deconfinement from AdS/CFT,''
  Nucl.\ Phys.\ B {\bf 896}, 738 (2015)
  [arXiv:1412.4964 [hep-th]].


\bibitem{Hartnoll:2008kx} 
  S.~A.~Hartnoll, C.~P.~Herzog and G.~T.~Horowitz,
  ``Holographic Superconductors,''
  JHEP {\bf 0812}, 015 (2008)
  [arXiv:0810.1563 [hep-th]].
	
\bibitem{Hartnoll:2008vx} 
  S.~A.~Hartnoll, C.~P.~Herzog and G.~T.~Horowitz,
  ``Building a Holographic Superconductor,''
  Phys.\ Rev.\ Lett.\  {\bf 101}, 031601 (2008)
  [arXiv:0803.3295 [hep-th]].

\bibitem{Hartnoll:2009sz} 
  S.~A.~Hartnoll,
  ``Lectures on holographic methods for condensed matter physics,''
  Class.\ Quant.\ Grav.\  {\bf 26}, 224002 (2009)
  [arXiv:0903.3246 [hep-th]].

\bibitem{Herzog:2009xv} 
  C.~P.~Herzog,
  ``Lectures on Holographic Superfluidity and Superconductivity,''
  J.\ Phys.\ A {\bf 42}, 343001 (2009)
  [arXiv:0904.1975 [hep-th]].
  
\bibitem{Sachdev:2010ch} 
  S.~Sachdev,
  ``Condensed Matter and AdS/CFT,''
  Lect.\ Notes Phys.\  {\bf 828}, 273 (2011)
  [arXiv:1002.2947 [hep-th]].

\bibitem{Li:2013fhw} 
  W.~J.~Li, Y.~Tian and H.~b.~Zhang,
  ``Periodically Driven Holographic Superconductor,''
  JHEP {\bf 1307}, 030 (2013)
  [arXiv:1305.1600 [hep-th]].

\bibitem{Natsuume:2013lfa} 
  M.~Natsuume and T.~Okamura,
  ``The enhanced holographic superconductor: is it possible?,''
  JHEP {\bf 1308}, 139 (2013)
  [arXiv:1307.6875 [hep-th]].

\bibitem{Oka2016} 
T. Oka and L. Bucciantini,
  Phys.\ Rev.\ B {\bf 94}, 155133 (2016).


\bibitem{Shirley:1965}
  J.H.~Shirley,
  Phys.\ Rev.\ B {\bf 138}, 979 (1965).

 \bibitem{Morimoto:2016} 
T.~Morimoto, N.~Nagaosa,
Scientific Reports {\bf 6}, 19853 (2016).


\bibitem{Kim:2011qh} 
  K.~-Y.~Kim, J.~P.~Shock and J.~Tarrio,
  ``The open string membrane paradigm with external electromagnetic fields,''
  JHEP {\bf 1106}, 017 (2011)
  [arXiv:1103.4581 [hep-th]].

\bibitem{Seiberg:1999vs} 
  N.~Seiberg and E.~Witten,
  ``String theory and noncommutative geometry,''
  JHEP {\bf 9909}, 032 (1999)
  [hep-th/9908142].

\bibitem{Gibbons:2000xe} 
  G.~W.~Gibbons and C.~A.~R.~Herdeiro,
  ``Born-Infeld theory and stringy causality,''
  Phys.\ Rev.\ D {\bf 63}, 064006 (2001)
  [hep-th/0008052].

\bibitem{Gibbons:2001ck} 
  G.~W.~Gibbons,
  ``Pulse propagation in Born-Infeld theory: The World volume equivalence principle and the Hagedorn - like equation of state of the Chaplygin gas,''
  Grav.\ Cosmol.\  {\bf 8}, 2 (2002)
  [hep-th/0104015].

\bibitem{Gibbons:2002tv} 
  G.~Gibbons, K.~Hashimoto and P.~Yi,
  ``Tachyon condensates, Carrollian contraction of Lorentz group, and fundamental strings,''
  JHEP {\bf 0209}, 061 (2002)
  [hep-th/0209034].

\bibitem{OkaAoki} 
  T.~Oka and H.~Aoki,
``All Optical Measurement Proposed for the Photovoltaic Hall Effect,''
J. Phys.: Conf. Ser. {\bf 334} 012060 (2011)
[arXiv:1007.5399].


\bibitem{Burkov} 
  A.~A.~Burkov and L.~Balents,
``Weyl Semimetal in a Topological Insulator Multilayer,''
Phys.\ Rev,\ Lett.\ {\bf 107}, 127205 (2011)
[arXiv:1105.5138].


\bibitem{soon} 
  K.~Hashimoto, S.~Kinoshita, K.~Murata and T.~Oka, {\it in preparation}.



\end{thebibliography}
\end{document}